\documentclass[aps,preprint,onecolumn,eqsecnum,nofootinbib]{revtex4}

\usepackage{epsfig}
\usepackage{graphicx}
\usepackage{dcolumn}
\usepackage{amsmath}
\usepackage{enumerate}
\usepackage{subfigure}

\def\gtap{\mathrel{ \rlap{\raise 0.511ex \hbox{$>$}}{\lower 0.511ex
   \hbox{$\sim$}}}} 
\def\ltap{\mathrel{ \rlap{\raise 0.511ex
    \hbox{$<$}}{\lower 0.511ex \hbox{$\sim$}}}} 
\newcommand{\vs}{\vspace{0.2truecm}}

\begin{document}

\vskip-6pt \hfill {FTUV-09-0226} 
\vskip-6pt \hfill {IPPP/09/13} 
\vskip-6pt \hfill {DCPT/09/26} 
\vskip-6pt \hfill {CFTP/09-17} \\

\title{A combined beta-beam and electron capture neutrino experiment}

\author{Jos\'{e} Bernab\'{e}u$^{1}$, Catalina Espinoza$^{1}$, \\
        Christopher Orme$^2$, Sergio Palomares-Ruiz$^{3}$ and Silvia
        Pascoli$^{2}$ \vspace{5mm}} 

\affiliation{$^{1}$ Universitat de Val\`encia and IFIC, E-46100,
        Val\`encia, Spain} 
\affiliation{$^{2}$ IPPP, Department of Physics, Durham University,
        Durham, DH1 3LE, United Kingdom} 
\affiliation{$^{3}$ Centro de F\'{\i}sica Te\'orica de
        Part\'{\i}culas, Instituto Superior T\'ecnico, 1049-001
        Lisboa, Portugal \vspace{5mm}}

\begin{abstract}
The next generation of long baseline neutrino experiments will aim 
at determining the value of the unknown mixing angle, $\theta_{13}$,
the type of neutrino mass hierarchy and the presence of CP-violation 
in the lepton sector. Beta-beams and electron capture experiments have
been studied as viable candidates for long baseline experiments. They
use a very clean electron neutrino beam from the $\beta$-decays or
electron capture decays of boosted ions. In the present article we
consider an hybrid setup which combines a beta-beam with an electron
capture beam by using boosted Ytterbium ions. We study the sensitivity
to the CP-violating phase $\delta$ and the $\theta_{13}$ angle, the
CP-discovery potential and the reach to determine the type of neutrino
mass hierarchy for this type of long baseline experiment. The analysis
is performed for different neutrino beam energies and baselines.
Finally, we also discuss how the results would change if a better
knowledge of some of the assumed parameters was achieved by the time
this experiment could take place.
\end{abstract}

\maketitle


\section{Introduction}

In the past decade, atmospheric~\cite{SKatm,atm},
solar~\cite{sol,SKsolar,SNO}, reactor~\cite{CHOOZ,PaloVerde,KamLAND} 
and long-baseline accelerator~\cite{K2K,MINOS} neutrino  experiments
have provided compelling evidence for the phenomenon of neutrino
oscillations. This has reshaped our understanding of the properties of
elementary particles as it implies that neutrinos have mass and mix.
The combined data can be described by two mass squared differences,
$\Delta m_{31}^{2}$ and $\Delta m_{21}^{2}$, where $\Delta
m_{ji}^{2}=m_{j}^{2}-m_{i}^{2}$, whose current best fit values are
$|\Delta m_{31}^{2}|=2.4 \times 10^{-3} \ \mathrm{eV}^2$ and   
$\Delta m_{21}^{2}=7.65 \times 10^{-5} \ \mathrm{eV}^2$~\cite{STV08}. 
The two mixing angles $\theta_{12}$ and $\theta_{23}$ drive the solar
and KamLAND, and atmospheric and MINOS neutrino oscillations,
respectively, and are measured to be $\sin^2 \theta_{12} = 0.304$ and 
$\sin^2 \theta_{23} = 0.50$~\cite{STV08}. The third mixing angle,
$\theta_{13}$, is yet undetermined but is known to be small or
zero. With available data, $\theta_{13}$ is constrained to
be~\cite{STV08}
\begin{equation}
\sin^{2}\theta_{13} < 0.040 \: (0.056) \quad \mbox{at} \quad  2\sigma
\: (3\sigma) ~.
\end{equation}
It is interesting to note that very recently a first hint in favour
of $\theta_{13} \neq0$ has been found~\cite{Fogli:2008jx} in a
combined analysis of atmospheric, solar and long-baseline reactor
neutrino data, with: 
\begin{equation}
\sin^2\theta_{13} = 0.016 \pm  0.010 \qquad \mbox{at 1$\sigma$} ~,
\end{equation}
implying a preference for $\theta_{13} > 0$ at 90\%CL. A different
analysis~\cite{Ge:2008sj} confirms the hint for $\theta_{13}\neq0$ at
1.5~$\sigma$ from the analysis of solar and KamLAND data owing to the
latest SNO results, but not the one from the atmospheric data.
 
Although the experimental progress in neutrino physics over the last
decade has been conspicuous, many of the fundamental questions
surrounding neutrinos still need to be addressed. Understanding of the 
physics beyond the Standard Model responsible for neutrino masses and
mixing requires knowledge of the nature of neutrinos (whether Dirac
or Majorana particles), the neutrino mass ordering (normal or
inverted), the absolute neutrino mass scale, the value of the unknown  
mixing angle $\theta_{13}$, and whether CP-symmetry is violated in the
lepton sector. It will also be necessary to improve the precision on
the known parameters, in particular to measure any deviation from
maximal $\theta_{23}$ and, if so, to determine its octant.

Some of the issues above will be addressed by a future program of
neutrino oscillation experiments~\cite{ISS,Euronu}. In particular,
long baseline experiments using conventional beams~\cite{MINOS} and
nuclear reactors~\cite{freactors} will be the first to explore
$\theta_{13}$ below the current limit and maybe confirm the hint for
$\theta_{13} \neq 0$~\cite{Fogli:2008jx}. If $\theta_{13}$ is close to
the present bound imposed by the running and near future experiments,
the next generation of superbeams~\cite{T2K,newNOvA}, an extension of
a conventional beam with an upgrade in intensity and detector size,
and wide-band beams~\cite{Barger:2007jq} will probe CP-violation and,
for sufficiently long baseline, the neutrino mass hierarchy. For small
values of $\theta_{13}$ or, if  $\theta_{13}$ is large but a better
precision on the neutrino parameters needs to be achieved, the
community must turn to the novel concepts of the neutrino
factory~\cite{nufact,nufactlow} or beta-beam~\cite{zucchelli,mauro}.
Whereas conventional beams sourced from pion decays
have an intrinsic contamination of electron neutrino at the $\sim$ 1\%
level (owing to kaons in the beam), neutrino factories and beta-beams
will have clean sources from highly accelerated muons and ions,
respectively, producing a well-collimated beam. In a neutrino factory,
muons (antimuons) are produced, cooled and accelerated to a high boost
before being stored in a decay ring. The subsequent decay sources a
muon neutrino (muon antineutrino) and electron antineutrino (electron
neutrino) which are aimed at magnetised detectors located a very long
distance from the source. The use of magnetised detectors is necessary
to separate the `right muon' disappearance signal from the `wrong
muon' appearance signal, which is sensitive to matter effects and
CP-violation. A beta-beam will exploit accelerated ions that
$\beta$-decay sourcing a clean, collimated, electron neutrino beam.
Magnetised detectors will not be necessary in this case, the only
requirement being possession of good muon identification to detect the
appearance channels. Therefore, water \v{C}erenkov (WC), totally
active scintillator, liquid argon detectors and non-magnetised iron
calorimeters could be used, depending on the peak energy.

The determination of the oscillation parameters is severely affected
by degeneracies~\cite{deg1,deg2,deg3,deg4,deg5}; the possibility that
different sets of the unknown parameters $(\mbox{sgn}(\Delta
m_{31}^{2}),\delta, \theta_{13}, \theta_{23}$ octant) can provide an 
equally good fit to the probability for neutrino and antineutrino
oscillations, for fixed baselines and energy. Therefore, a high
precision measurement of the appearance probabilities is not
sufficient to discriminate the various allowed solutions. In order to
weaken or resolve this issue, various strategies have been put forward:
exploiting the energy dependence of the signal in the same experiment 
~\cite{Barger:2007jq,singleion}, using reactor neutrino experiments
with an intermediate baseline~\cite{reactorinter}, combining different
long baseline experiments~\cite{otherexp}, adding the information on
$\theta_{13}$ from reactor experiments~\cite{reactorlong}, or using
more than one baseline for the same
beam~\cite{silver,MN97,BMW02off,SN,twodetect,CDFL07}. In addition,  
$\theta_{13}$ controls the Earth matter effects in multi-GeV
atmospheric~\cite{atmmatter1,mantle,core,atmmatter2,atmmatter3} and in
supernova neutrino oscillations~\cite{SNastro} (see also
Ref.~\cite{DDM08}). These might provide useful information on the type
of neutrino mass hierarchy and $\theta_{13}$; the magnitude of the
T-violating and CP-violating terms in neutrino oscillation
probabilities is directly proportional to
$\sin\theta_{13}$~\cite{CPT,BB00}. 

In beta-beam experiments, the energy dependence of the signal is
typically used to extract information on the mass hierarchy and
CP-violation. Matter effects increase with baseline and energy
suggesting that setups with baselines $>$ 600 km are
necessary~\cite{CDFL07,betabeamhigh,bblindner,alternating,betaCERNupgrade,bblhc,singleion}
for the determination of the type of neutrino mass ordering. Such
strategies would make use of a proposed upgrade to the CERN Super
Proton Synchrotron (SPS) which would equip the accelerator with fast 
superconducting magnets allowing high boosts and fast ramps. The
latter are important to reduce the loss of ions through decay in the
acceleration stage. A sister approach to the beta-beam is to use the
neutrinos sourced from ions that decay mainly through electron
capture~\cite{EC1,EC2,Sato,RS06}. If the
electron capture decay is dominated by a single channel, then a 
monoenergetic electron neutrino beam can be produced this way. In this
case, all the beam intensity can be concentrated at the appropriate
energy to get the best sensitivity to the oscillation parameters. In
order to disentangle the CP violating phase with neutrino beams only,
one makes use of the different energy dependence of the CP-even and
CP-odd terms in the appearance probability~\cite{BB00}. Electron
capture competes with $\beta^{+}$-decay when the $Q_{\rm EC}$-value $> 
2m_{e}$, $m_{e}$ being the electron mass. With the ions identified
in~\cite{EC1}, the use of an upgraded SPS or the
Tevatron~\footnote{Note that the present Tevatron configuration does 
  not ramp fast enough, resulting in a high loss of ions, so this
  might not be a very realistic experimental setup, at least in the
  present configuration.} is necessary to source baselines in excess
of CERN-Frejus (130 km). In this paper, we discuss a hybrid of these
two approaches.~\footnote{Note that the use of this hybrid approach
  was first mentioned in Ref.~\cite{Sato}, although the proposal,
  unlike the present case, was to use long-lived ions. In addition,
  the phenomenology of this approach was not studied.} By selecting a 
nuclide with $Q_{\rm EC}\sim 4$~MeV, we can make use of neutrinos from
an electron capture $\delta$-spike and $\beta^{+}$ continuous spectrum
simultaneously. Assuming a detector with low energy threshold, the use
of such ions allows one to exploit the information from the first and
second oscillation maxima with a single beam, in a similar way to the
approach used in Ref.~\cite{singleion}. There the spectral information
was used to remove some of the degeneracies and reach physics
sensitivities comparable to the scenarios with a neutrino and
antineutrino beam that are often presented in the literature. The use
of the hybrid approach we propose makes it possible to use a
monochromatic beam at higher energies and a beta-beam at lower
energies. The need for good neutrino energy resolution at the higher
energies will therefore be less crucial than for high-$\gamma$
beta-beam scenarios.
 
The paper is organised as follows. In Sec.~\ref{sec:concept}, we
introduce the hybrid idea and show two possible ions which could be
used. In Sec.~\ref{sec:setups} we present the different set-ups
studied, discussing the choice of boost factor, baseline and various
detector options. The description of the simulation and analysis is
presented in Sec.~\ref{sec:analysis} and the results in
Sec.~\ref{sec:results}. The final discussion and conclusions are drawn
in Secs.~\ref{sec:discussion} and~\ref{sec:conclusions}.


\section{The beta-beam and electron capture combination} 
\label{sec:concept}

In this section we introduce the idea of the beta-beam and electron
capture hybrid approach. We present the spectra of the two branches,
their ratio and discuss two nuclides which have desirable properties. 

The beta-beam is a proposal, originally put forward by
P.~Zucchelli~\cite{zucchelli}, to accelerate and then store
$\beta$-emitting ions, which subsequently decay to produce a well
collimated, uncontaminated, electron neutrino (or antineutrino)
beam. The high luminosities required to achieve a useful physics reach
point towards ions with small proton numbers to minimise space charge
and half-lives $\sim$ 1~second to reduce ion losses during the
acceleration stage whilst maintaining a large number of useful decays
per year. The most promising candidate ions are $^{18}$Ne and $^{8}$B
for neutrinos, and $^{6}$He and $^{8}$Li for antineutrinos. A variant
on the beta-beam idea is the use of electron capture to produce
monoenergetic neutrino beams. Electron capture is the process in which
an atomic electron is captured by a bound proton of the ion $A(Z,N)$
leading to a nuclear state of the same atomic number $A$, but with the
exchange of the proton by a neutron and the emission of an electron
neutrino, 
\begin{equation}
A(Z,N) + e^- \rightarrow A(Z-1,N+1) + \nu_e ~.
\end{equation}
The idea of using this process in neutrino experiments was
independently discussed in Refs.~\cite{EC1,Sato}. In Ref.~\cite{RS06},
ions with low $Q_{\rm EC}$-value and long half-life, such as
$^{110}$Sn, were proposed to be accelerated to very high boosts with
the LHC. Baselines of 250~km and 600~km were considered with the
spectral information coming from the position of the events in the
detector. Sensitivities comparable to a Neutrino Factory were obtained
for a single boost. However, in order for electron capture machines to
become operational, nuclei with shorter half-life are required. The
recent discovery of nuclei far from the stability line with
kinematically accessible super-allowed spin-isospin transitions to 
giant Gamow-Teller resonances (see, for example, Ref.~\cite{RG09})
opens up such a possibility. The rare-Earth nuclei above $^{146}$Gd
have a short enough half-life to allow electron capture processes in
the decay ring, in contrast to fully-stripped long-lived
ions~\cite{Sato,RS06}. This was the alternative put forward in
Ref.~\cite{EC1} where the use of short-lived ions with $Q_{\rm
  EC}$-values around 1-4 MeV was proposed. Machines such as the SPS,
an upgraded SPS and the Tevatron could then be used for the
acceleration. The ion $^{150}$Dy, with $Q_{\rm EC}$-value 1.4~MeV, was
investigated for the CERN-Frejus (130 km) and CERN-Canfranc (650 km)
baselines and different boost factors. It was found to have very good
physics reach~\cite{EC1,EC2}. Owing to the monochromatic nature of the
beam, multiple boosts are necessary to resolve the intrinsic
degeneracy in this case.

In the following, we demonstrate how the flux for the electron
capture/beta-beam can be built up by discussing them separately and
comparing branching ratios. Let the mass difference between the parent
and the daughter nuclei, $\Delta M_A^{\beta^+} = M_A(Z,N) -
M_A(Z-1,N+1)$, include the mass and the binding energy of an atomic
electron as well. For electron capture, the maximum kinetic energy
release is thus given by $Q_{\rm EC} =  \Delta M_A^{\beta^+}$. 
For $\beta^+$-decay, however, the final nucleus has an excess electron
since a positron is produced. The maximum kinetic energy release is
thus given by $Q_{\beta^+} = \Delta M_A^{\beta^+} - 2 \, m_{e}$.
Clearly for $(\Delta M_A^{\beta^+} = ) \, Q_{\rm EC} <  2m_{e}$,
electron capture is the only allowed process for a proton-rich
nucleus. For $Q_{\rm EC} > 2m_{e}$, electron capture and positron
emission compete, their branching ratios dependent on $Q_{\rm EC}$. If
decay through $\alpha$ emission is also allowed, it is important that
this has a relatively low $Q$-value so as not to be the dominant
channel~\footnote{The $\alpha$ decay branching ratio is strongly
  dependent on the $Q_{\rm EC}$-value. For low $Q_{\rm EC}$, the
  $\alpha$ decay probability is sufficiently long as to allow the weak
  decay modes to be the main channels.}. For a number of useful ion
decays per year $N_{\rm ions}$, the electron capture neutrino flux is
given by~\cite{EC1,EC2}
\begin{equation}
\frac{d\Phi^{\rm lab}_{\rm EC}}{d\Omega
  dE_\nu}=\frac{\Gamma}{\Gamma_{\rm tot}}\,\frac{N_{\rm ions}}{\pi
  L^{2}}\,\gamma^{2}\,\delta(E_\nu-2\gamma E_0^{\rm EC})
\end{equation}
for each decay channel. Here, $L$ is the baseline, $\gamma$ is the
Lorentz boost, $E_0^{\rm EC}$ ($= Q_{\rm EC})$ is the neutrino energy
in the ion rest frame and $E_\nu$ is the neutrino energy in the lab
frame.

The flux for the $\beta$-spectrum is found in the usual way. In the
rest frame of the ion, the electron neutrino flux is proportional to 
\begin{equation}
\frac{d\Phi^{\rm rf}_\beta}{d\cos\theta dE_{\rm rf}} \sim
E_{\rm rf}^2 (E_0^\beta-E_{\rm rf})\sqrt{(E_{\rm rf} - E_0^\beta)^{2}
  - m_{e}^{2}}
~. 
\label{E:flux}
\end{equation}
Here, $E_0^\beta$ ($= Q_{\beta^+} + m_e = Q_{\rm EC} - m_e$) is the
total end-point energy of the decay. The neutrino flux per solid angle
at the detector located at distance $L$ from the source after boost
$\gamma$ is~\cite{betabeamhigh}  
\begin{equation}
\left.\frac{d\Phi^{\rm lab}_\beta}{d\Omega dy}\right|_{\theta\simeq 0}
\simeq \frac{N_{\rm ions}}{\pi  L^{2}}
\frac{\gamma^{2}}{g(y_{e})}y^{2}(1-y)\sqrt{(1-y)^{2}-y_{e}^{2}} ~,
\label{E:Bflux}
\end{equation}
where $0 \leq y=\frac{E_{\nu}}{2\gamma E_0^\beta}\leq 1-y_{e}$,
$y_{e}=m_{e}/E_0^\beta$, and 
\begin{equation}
g(y_{e})\equiv \frac{1}{60}\left\{
\sqrt{1-y_{e}^{2}}(2-9y_{e}^{2}-8y_{e}^{4})+15y_{e}^{4}
\log\left[\frac{y_{e}}{1-\sqrt{1-y_{e}^{2}}}\right] \right\}. 
\end{equation}
Similarly to the case of electron capture, a neutrino with energy
$E_{\rm rf}$ in the rest frame will have a corresponding energy
$E_{\nu} = 2\gamma E_{\rm rf}$ in the laboratory frame along the
$\theta =0^{\circ}$ axis.  

All the known nuclear structure information on the $A=148$ and $A=156$
nuclides has been reviewed in Ref.~\cite{Bhat} and Ref.~\cite{Reich},
respectively, where the information obtained in various reaction and
decay experiments is presented, together with adopted level schemes.
Currently, a systematic study of electron capture decays in the region
of $^{146}$Gd, relevant for monoenergetic neutrino beams, is being
carried out~\cite{Ybcom}. Here, we consider two nuclides,
$^{156}_{70}$Yb and $^{148m}_{65}$Tb, that decay through electron
capture and $\beta^{+}$-decay with similar branching ratios whose
lifetimes are not too long or too short. Their decays are summarised
in Tables~\ref{T:decay_Yb} and~\ref{T:decay_Tb}. Ytterbium is a
nuclide $^{156}_{70}$Yb with spin-parity $0^+$, which decays 90\% via
electron capture plus $\beta^+$-decay~\cite{Reich}, with 38\% via
electron capture and 52\% via $\beta^+$-decay~\cite{Ybcom}. The
remaining 10\% goes into alpha particles and a different final
state. This relatively small branching ratio into alphas helps the
nuclide to have a short enough half-life, 26.1~seconds. It is
important to note that this electron capture-$\beta^+$-decay
transition has only one possible daughter state with spin-parity
$1^+$, i.e., it is a Gamow-Teller transition into an excited state of
Thulium, $^{156}_{69}$Tm$^*$. The transition $Q_{\rm EC}$-value
is~\footnote{$Q_{\rm EC}$-values are typically calculated between
  ground states unless stated otherwise.} $Q_{\rm EC}$-value = 3.58
MeV. However, the excitation energy of the final nuclear state
(0.12~MeV) needs to be taken into account and thus, the effective
$Q_{\rm EC}$-value (difference in the total kinetic energies of the
system after and before the decay) is 3.46~MeV~\cite{Reich}. The
electron capture energy of $\sim 4$ MeV is well suited to the
intermediate-baselines of Europe and the USA with the available
technology, or those available with future upgrades. On the other 
hand, the $^{148m}_{65}$Tb isomer with spin-parity $9^+$ has a $Q_{\rm
  EC}$-value of 5.77~MeV~\cite{Bhat,Tb}. Although the decay to the
ground state of $^{148}_{64}$Gd is highly forbidden, the presence of a
Gamow-Teller resonance allows the decay into an excited state with
effective $Q$-value 3.07~MeV~\cite{GSI}. This nuclide is longer lived  
than $^{156}_{70}$Yb (with a half-life of 2.2 minutes) and will require
slightly higher boosts. It is still well suited to intermediate
baselines. However, the dominance of the electron capture over the
$\beta^{+}$-decay channel makes this nuclide less desirable. The count
rate will be dominated by the single energy of the electron capture
which provides insufficient information to obtain the good
sensitivities aspired to by future long baseline experiments. It was
shown in Refs.~\cite{EC1,EC2} that two runs with different boosts are
necessary for an exclusive or dominant electron capture channel to
break the intrinsic degeneracy and achieve good CP-violation
discovery. Hence, in what follows we will study this hybrid approach
focusing on $^{156}$Yb.

\begin{table}
\begin{center}
\begin{tabular}{|c|c|c|c|}
\hline
\quad Decay \quad & \quad Daughter \quad & \quad Neutrino energy (MeV)
\quad & \quad BR \quad \\  
&&& \\
\hline
\hline
$\beta^{+}$ & $^{156}_{69}$Tm$^{*}$ & 2.44 {\rm (endpoint)} & 52\% \\
EC & $^{156}_{69}$Tm$^{*}$ & 3.46 \hspace{1.77cm} & 38\%\\
$\alpha$ & $^{152}_{68}$Er & 4.81 \hspace{1.77cm} & 10\% \\
\hline
\end{tabular}
\end{center}
\caption{Decay summary for $^{156}_{70}$Yb. The $Q_{\rm EC}$-value for
  the transition between ground states is 3.58 MeV and taking into
  account the excitation energy of the final nuclear state (0.12~MeV),
  the effective $Q^{\rm eff}_{\rm EC}$-value is
  3.46~MeV~\cite{Reich,Ybcom}.}
\label{T:decay_Yb}
\end{table}

\begin{table}
\begin{center}
\begin{tabular}{|c|c|c|c|}
\hline
\quad Decay \quad & \quad Daughter \quad & \quad Neutrino energy (MeV)
\quad & \quad BR \quad \\ 
& & & \\
\hline
\hline
$\beta^{+}$ & $^{148}_{64}$Gd$^{*}$ & 2.05 {\rm (endpoint)} & 32\% \\
EC & $^{148}_{64}$Gd$^{*}$ & 3.07 \hspace{1.77cm} & 68\%\\
\hline
\end{tabular}
\end{center}
\caption{Decay summary for $^{148\, m}_{65}$Tb. The $Q_{\rm EC}$-value
  for the transition between ground states is 5.77~MeV and the
  effective $Q^{\rm eff}_{\rm EC}$-value to the excited state is
  3.07~MeV~\cite{Tb,Bhat,GSI}.} 
\label{T:decay_Tb} 
\end{table}


\section{Experimental setups}
\label{sec:setups}

In this section we consider different boosts, baselines and
detectors. We first discuss the available (or possible future)
accelerator technology and identify the possible boost factors in
combination with different baselines. We then discuss the main
characteristics of the detectors considered in the analysis.

\subsection{Choice of $\gamma$ and baseline}

\begin{figure}
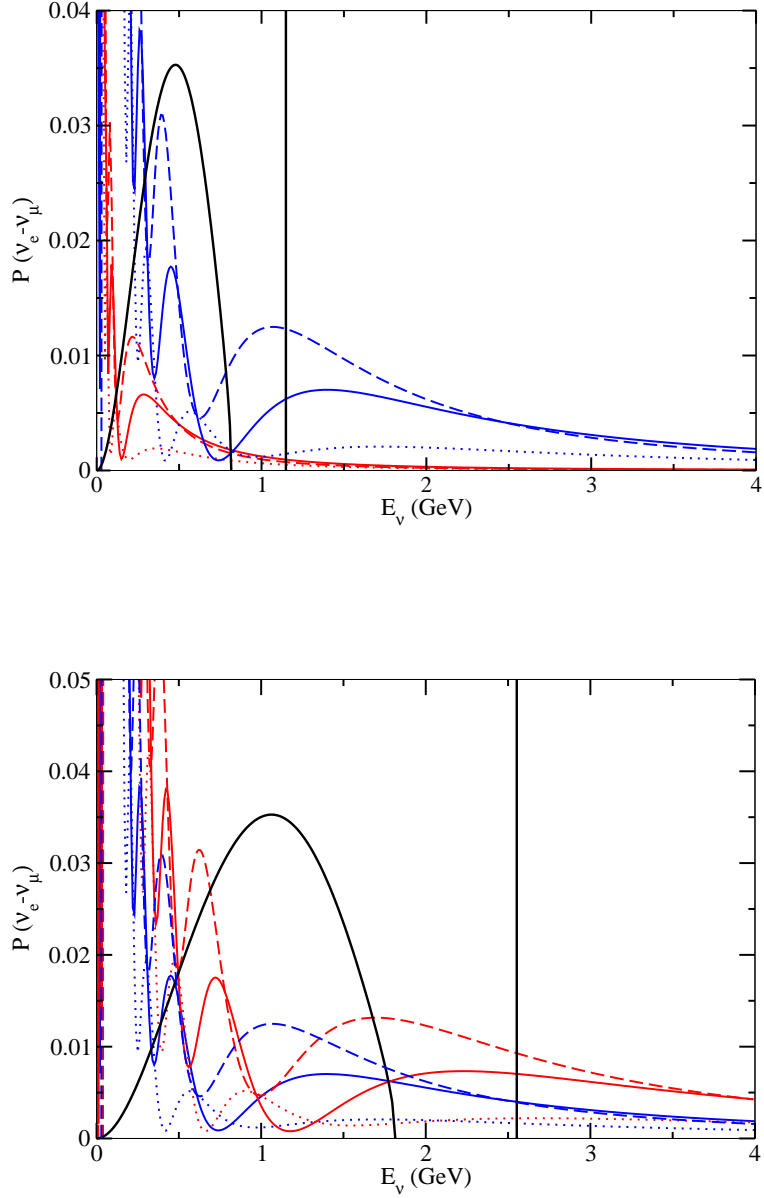

\begin{center}
\subfigure{\includegraphics[width=10cm,height=7cm]{G166.eps}}\\
\vspace{1.5cm}
\subfigure{\includegraphics[width=10cm,height=7cm]{G369.eps}}
\end{center}
\caption{{\it Top panel}: $\nu_{e}\rightarrow \nu_{\mu}$ appearance
  probabilities for the CERN-Frejus (130 km) and CERN-Canfranc (650 km)
  baselines. The unoscillated $\nu_{e}$ flux in the laboratory frame
  is shown for $^{156}$Yb given a boost $\gamma=166$ in arbitrary
  units. {\it Bottom panel}: $\nu_{e}\rightarrow \nu_{\mu}$ appearance
  probabilities for the CERN-Canfranc (650 km) and CERN-Boulby (1050
  km) baselines. The flux from a boost $\gamma=369$ is shown in
  arbitrary units. In both cases, the blue lines correspond to
  CERN-Canfranc; the red being CERN-Frejus (top panel) and CERN-Boubly
  (bottom panel). The solid lines correspond to $\delta=0^{\circ}$,
  dashed $\delta=90^{\circ}$ and dotted   $\delta=-90^{\circ}$. The
  value $\sin^2 2\theta_{13} = 0.01$ was taken for all curves.} 
\label{F:ECprobs}
\end{figure}

In this paper we consider the use of a neutrino beam sourced from
boosted $^{156}$Yb ions directed along a single baseline. As described
above, both the electron capture and $\beta^{+}$-decay channels are to
an excited state of $^{156}$Tm with a $Q_{\rm EC}$-value of
3.46~MeV. In order to fully exploit the electron capture decay mode,
the nuclides cannot be fully stripped; at least 16 electrons being
left on the ion~\cite{Mats}. The maximum boost, $\gamma_{\rm max}$,
available is thus
\begin{equation}
\gamma_{\rm max}=\frac{E_{\rm acc}}{m_{p}}\frac{Z-16}{A} ~,
\end{equation}
where $m_{p}$ is the mass of the proton and $E_{acc}$ is the maximum
energy accessible with the accelerator. Current and future accelerator
facilities would be an ideal production environment. In this analysis,
we consider the maximum boosts available from the current SPS and
upgraded SPS (see Table~\ref{T:maxboosts}) for the following baselines:

\begin{enumerate}
\item \textbf{Boost $\gamma=166$ with current SPS}
\begin{itemize}
\item CERN-Frejus (130 km)
\item CERN-Canfranc (650 km)
\end{itemize}
\item \textbf{Boost $\gamma=369$ with an upgraded SPS}
\begin{itemize}
\item CERN-Canfranc (650 km)
\item CERN-Boulby (1050 km)
\end{itemize}
\end{enumerate} 
\begin{table}
\begin{center}
\begin{tabular}{|c|c|c|c|}
\hline
\;Machine\; &\; $\gamma_{\rm max} $\; &\; $2 \gamma_{\rm max} Q^{\rm
  eff}_{\rm EC}$ (GeV)\; &\; $2 \gamma_{\rm max} Q^{\rm
  eff}_{\beta^{+}} $ (GeV)\; \\
\hline
\hline
SPS & 166 & 1.15 & 0.81\\
\quad Upgraded SPS \quad & 369 & 2.55 & 1.80 \\
\hline
\end{tabular}
\end{center}
\caption{Maximum boosts and neutrino endpoint energies for $^{156}$Yb
  available for the current SPS setup and a proposed 1 GeV upgraded
  SPS.}
\label{T:maxboosts}
\end{table}
With the current magnetic rigidity of the SPS,  the electron capture
spike can be placed on first oscillation for the CERN-Canfranc
baseline (650 km) with the beta-beam spectrum peaking around the
second oscillation maximum (see Fig.~\ref{F:ECprobs}).

A detector with a low energy threshold is necessary to exploit the
oscillatory structure of the appearance probability. The second option
would make use of the upgrades to the CERN accelerator facilities
necessary for suggested LHC upgrades. With a 1 TeV SPS, the electron
capture beam could be placed at first oscillation maximum, or on the
probability tail, for the CERN-Boulby baseline (1050 km) or, using the
Tevatron, for the FNAL-Homestake baseline (1280 km). With an energy
threshold of 250 MeV, these setups could exploit the information at
second oscillation maximum to resolve some of the degeneracies in an
approach analogous to Ref.~\cite{singleion}, where the boost for
$^{18}$Ne was chosen so that the boosted spectrum covered both first
and second maximum. Owing to the spectral nature of the decay, in the
analysis in Ref.~\cite{singleion} it was difficult to determine to
what extent the highest energies contribute to the overall sensitivity
of the setup. With the electron capture beta-beam combination we can
investigate this issue as it is possible to place the electron capture
spike on first oscillation maximum or on its tail whilst the beta-beam
spectrum has minimal coverage of that energy range. In our analysis we
perform ($\theta_{13},\delta$) sensitivity contours through the
consideration of each decay channel separately and their
combination. This allows us to investigate the importance of the
contribution of each decay channel and evaluate if the importance of
the low and high energy contributions to the overall sensitivity.

\subsection{Choice of detector}

In a beta-beam, one aims to exploit the $\nu_{e} \rightarrow
\nu_{\mu}$ channel; a detector with excellent muon identification 
capabilities and efficient neutral current background rejection is
therefore required. For energies below $\sim$1~GeV, water-\v{C}erenkov
detectors are typically chosen with the muons identified through
the use of quasi-elastic events (QE). Efficient reduction of neutral
current events and (subdominant) pions is through the identification
of the decay process. For higher energies, the number of QE events
drops sharply where the deep inelastic scattering (DIS) component
dominates the cross section. Water-\v{C}erenkov detectors are usually
not the best choice for the higher energies owing to high backgrounds
and poor neutrino energy resolution. For large boosts, the 50~kton
class of detectors such as liquid argon (LAr), using time projection
chamber techniques, or total active scintillator detector (TASD),
based on tracking calorimeter principles, are usually considered. In
addition to QE events, these technologies also measure the energy
deposited through the hadronic channels and DIS events are in
principle also distinguishable. Their main disadvantage is their size
which is far smaller than the next generation water-\v{C}erenkov
detectors discussed in the literature which typically have fiducial
masses in the megaton scale.

For a pure electron capture machine, the choice of detector technology
does not depend on its energy reconstruction capabilities. In this
case, the neutrino energy is given by the choice of ion and 
boost factor leaving no need to reconstruct the neutrino energy in the
detector. For the hybrid approach we consider, it is possible, in
principle to separate the energy of the line spectrum from the
continuous spectrum. Suppose we identify an event and classify it as
being a QE elastic event with energy $E_{\nu}({\rm QE})$, then it must
be the case that the true energy $E_{\nu}^{{\rm true}} \geq
E_{\nu}({\rm QE})$. Thus, if one measures $E_{\nu}({\rm QE}) > 2
\gamma E_{0}^{\beta}$, then this event must be attributed to the
electron capture flux and hence, it is not necessary to reconstruct
more precisely the true neutrino energy. The separation between the
energy of the electron capture spike and the end-point energy of the
beta-spectrum is $2 m_e \gamma$. This should render the distinction
between the electron capture and the beta-beam fluxes possible. We
will assume this throughout the paper.

In the analysis, we will follow two strategies regarding the detector
type. We consider a generic detector technology, which could be LAr or
TASD, with a fiducial mass of 50~kton and assume that the neutrino
spectral information can be extracted from the charged current events. 
On the other hand we also consider a 0.5~Mton (fiducial)
water-\v{C}erenkov detector. In this case, following the prescription
described above, we assume the neutrino energy from beta-beam events
can only be reconstructed for QE events. However, we do include the
information from the inelastic events. As no spectral information is
possible for those events, we include them in a single bin. We will
take perfect efficiency for the 50~kton detector (which can be easily
scaled), and 70~\% efficiency for the water-\v{C}erenkov detector.


\section{Simulation and analysis}
\label{sec:analysis}

Based on the selected boost factors, baselines and type of detector
we have discussed in detail in the previous section,
in our analysis we will study and compare six different setups:

\begin{enumerate}
\item \textbf{50~kton detector (LAr or TASD) with $2 \times
  10^{18}$~ions/yr}
\begin{itemize}
\item Setup I: CERN-Frejus (130~km) and $\gamma = 166$
\item Setup II: CERN-Canfranc (650 km) and $\gamma = 166$
\item Setup III: CERN-Canfranc (650~km) and $\gamma = 369$
\item Setup IV: CERN-Boulby (1050~km) and $\gamma = 369$ 
\end{itemize}
\item \textbf{0.5~Mton water-\v{C}erenkov detector with
  $2 \times 10^{18}$~ions/yr} 
\begin{itemize}
\item Setup III-WC: CERN-Canfranc (650 km) and $\gamma = 369$
\item Setup IV-WC: CERN-Boulby (1050 km) and $\gamma = 369$
\end{itemize}
\end{enumerate} 
We will take a running time of 10~years for all the experimental
configurations considered. 

The number of events is computed for each energy bin $i$, given by  
\begin{equation}
n_i = {\cal{N}} \, \int dE \, \Phi(E) \, P(E) \, \sigma(E) \, K_i(E)
~,
\label{eq:nevents}
\end{equation}
where $\cal{N}$ is a constant which takes into account the efficiency,
the mass of the detector and the running time, $\Phi(E)$ is the
neutrino flux spectrum at the detector, $P(E)$ is the probability
function, $\sigma(E)$ represents the  total, QE or DIS cross section
(as described in the previous section) and $K_i(E)$ is the energy
smearing kernel for the $i$th bin for which we take a Gaussian energy
resolution function with a constant width of 150~MeV.

In the experimental simulations performed in this study, our analysis
is based on the following $\chi^{2}$ definition 
\begin{equation}
\label{cata7}
\chi^2 = \sum_{i,j} (n^t_i - n^f_i) C^{-1}_{ij}(n^t_j - n^f_j) ~,
\end{equation}
where $n^f_i$ are the predicted (or fitted) number of events for a
certain oscillation hypothesis, and $n^t_i$ are the simulated ``data''
for the true values of the parameters. The covariance matrix $C$ given
by
\begin{equation}\label{cata8}
C_{ij} = \delta_{ij} (\delta n^t_i)^2 ~,
\end{equation}
where $(\delta n^t_i) = \sqrt{n^t_i + (f_\textrm{{\small{sys}}} \cdot
  n^t_i)^2}$, contains both statistical and a 2\% overall systematic
error ($f_\textrm{{\small{sys}}} = 0.02$). In addition, we assume an
intrinsic beam background of 0.1~\% of the unoscillated spectrum 
originating from neutral current pion production and muons
misidentified as electrons. In the energy range of interest, there are
about 30 atmospheric neutrino events per kton-year which could mimic a 
muon coming from a $\nu_{e}\rightarrow \nu_{\mu}$ oscillation. We take 
$10^{-3}$ as the accelerator duty factor so that this atmospheric
background amounts to 0.03~events per kton-year.

In all setups considered in this paper, for the beta-beam part we take
an energy threshold of 250 MeV and use 200 MeV bins above that value,
except for non-QE events in water-\v{C}erenkov detectors which are
grouped in a single bin. For the electron capture events we always
take a single bin. Unless otherwise stated, we impose restrictions on
certain subsets of the fitted parameters in order to account for
external information from other experiments. Thus, we introduce the
so-called {\it{priors}}. Hence, if we want to restrict some parameter
$\kappa$, we introduce the central value $\kappa_c$ of the prior and
the input error $\sigma_{\kappa}$, and the actual minimisation is
performed over the modified $\chi^2$ function
\begin{equation}\label{cata9}
\chi^2 \rightarrow \chi^2 + \frac{(\kappa -
  \kappa_c)^2}{\sigma_{\kappa}^2} ~.
\end{equation}
In this work, we set priors for the experimentally known oscillation
parameters, taking their best fit values as central
values~\cite{STV08} of the corresponding priors, and the half width of
one standard deviation of their best fit values as the corresponding
input errors. Specifically, $\Delta m_{21}^{2}$, $|\Delta m_{31}^{2}|$,
$\sin^2 \theta_{12}$ and $\sin^2 \theta_{23}$ have been given errors
3\%, 5\%, 7\% and 14\% respectively. We marginalise over $\theta_{13}$
and $\delta$ over their entire range.


\section{Results}
\label{sec:results}

In this section we present and discuss the results of our detailed
numerical analysis of the various setups. In order to understand some
of the features of these results, it is useful to consider an
analytical approximation for the oscillation probability which for
these energies and baselines is given by~\cite{nf6} (see also
Ref.~\cite{Akhmedov:2004ny}): 
\begin{equation}
\begin{array}{l}
P(\nu_e \rightarrow \nu_\mu, L)  \simeq  
\sin^2 \theta_{23} \, \sin^2 {2 \theta_{13} } \left(
\frac{\Delta_{13}}{A- \Delta_{13}} \right)^2
\sin^2 \left( \frac{(A - \Delta_{13}) L}{2} \right) \\ \hspace{5mm}
+ \cos \theta_{13} \sin {2 \theta_{13} } \sin {2 \theta_{23}} 
\sin {2 \theta_{12}} \ \frac{\Delta_{12}}{A} \frac{\Delta_{13}}{A -
  \Delta_{13}} 
\sin \left(\frac{A L}{2} \right) \sin \left( \frac{(A -
  \Delta_{13}) L}{2} \right)
 \cos \left(\frac{\Delta_{13} L}{2}+ \delta \right) \\ \hspace{5mm}
+ \cos^2 \theta_{23} \sin^2 {2 \theta_{12}} \left(
 \frac{\Delta_{12}}{A} \right)^2 \sin^2 \left( \frac{A L}{2} \right),
\end{array}
\label{eq:probappr}
\end{equation}
where $\Delta_{12} \equiv \Delta m^2_{21} /(2 E)$ and $\Delta_{13}
\equiv \Delta m^2_{31} /(2E)$. We use the constant density
approximation for the index of refraction in matter $A \equiv \sqrt{2}
G_{F} \bar{n}_e(L)$, with $\bar{n}_e(L)= 1/L \int_{0}^{L} n_e(L') dL'$
the average electron number density. We analyse the sensitivity to
$\theta_{13}$ and $\delta$ for all the setups, and discuss the
discovery reach for CP-violation and the type of neutrino mass
ordering.

\begin{figure}
\begin{center}
\hspace{-1.5cm}
\subfigure{\includegraphics[width=9.5cm]{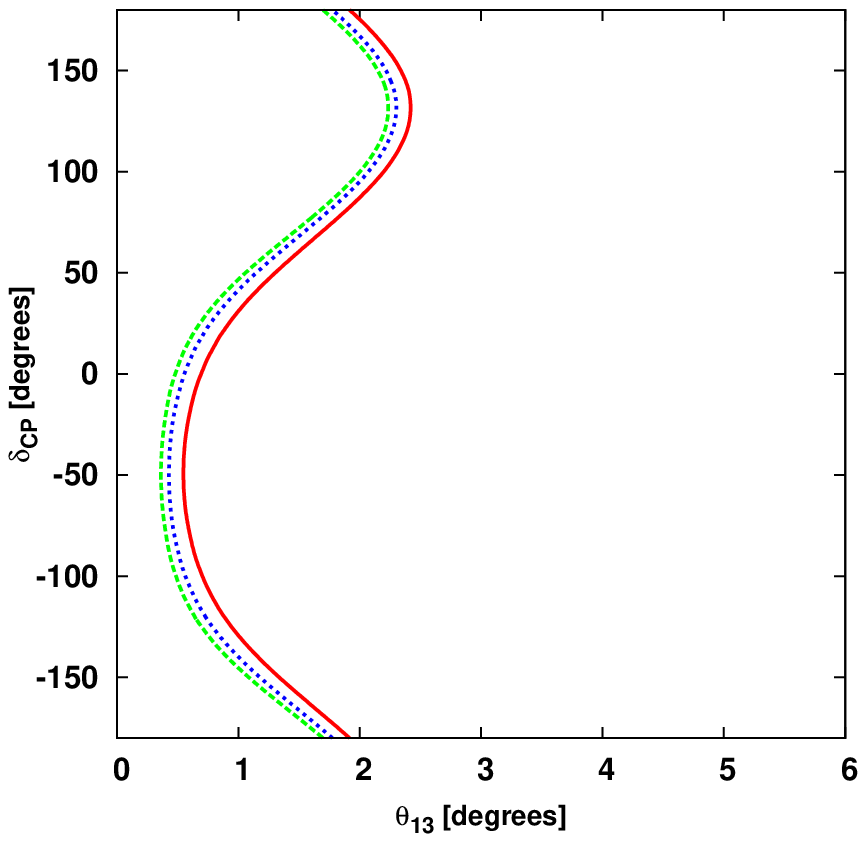}} \hspace{-2cm}
\subfigure{\includegraphics[width=9.5cm]{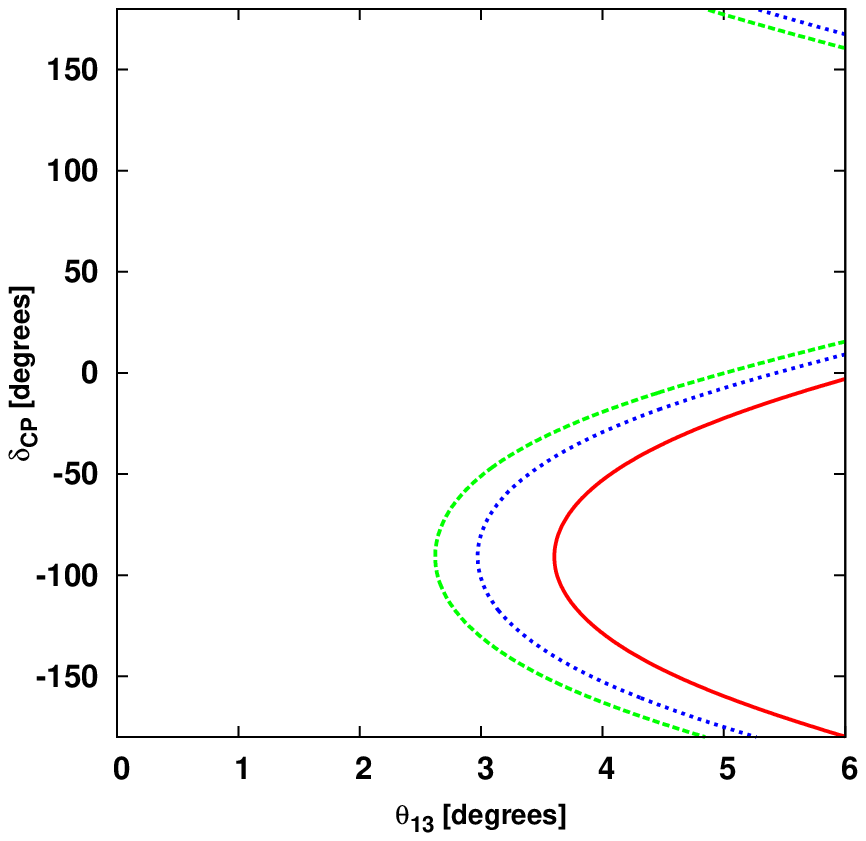}}\\
\hspace{-1.5cm}
\subfigure{\includegraphics[width=9.5cm]{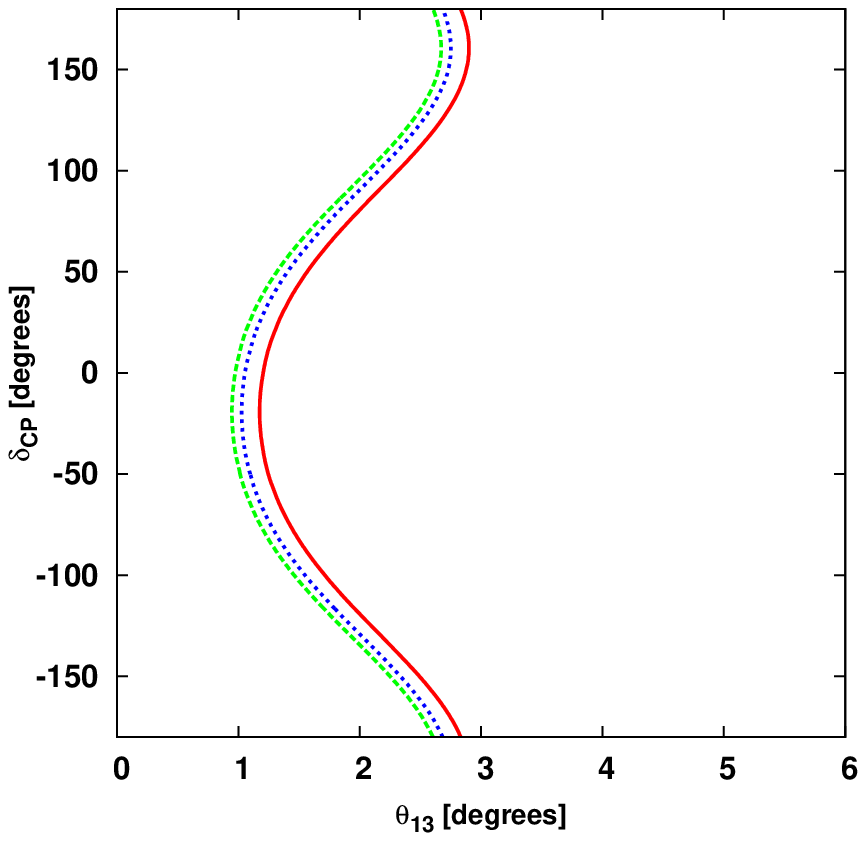}} \hspace{-2cm}
\subfigure{\includegraphics[width=9.5cm]{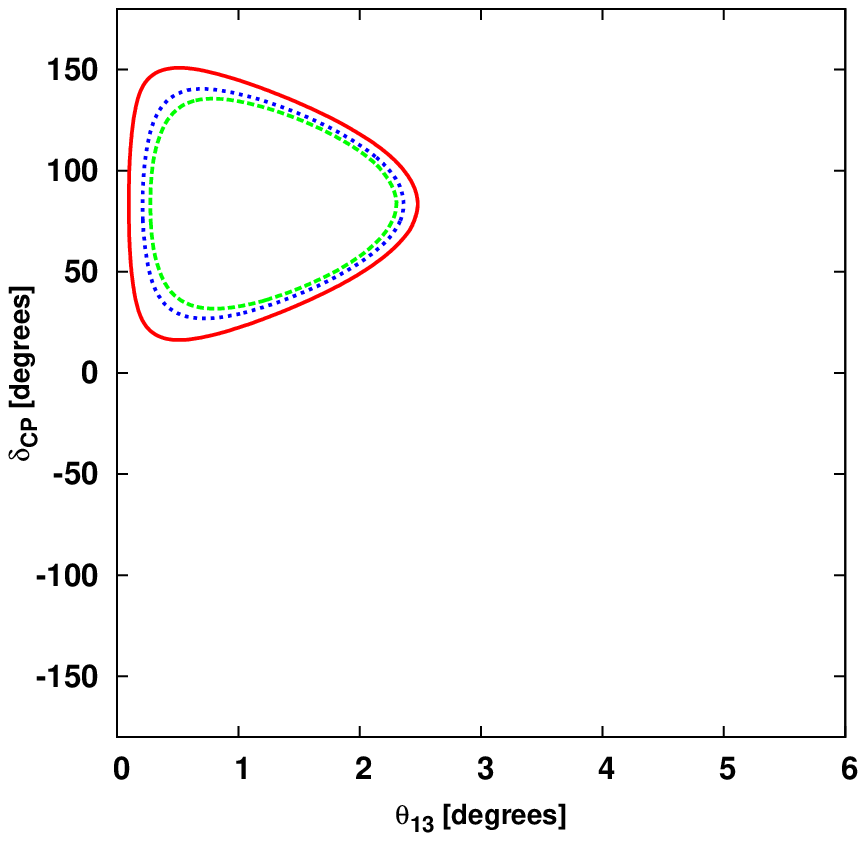}}\\
\hspace{-1.5cm}
\subfigure{\includegraphics[width=9.5cm]{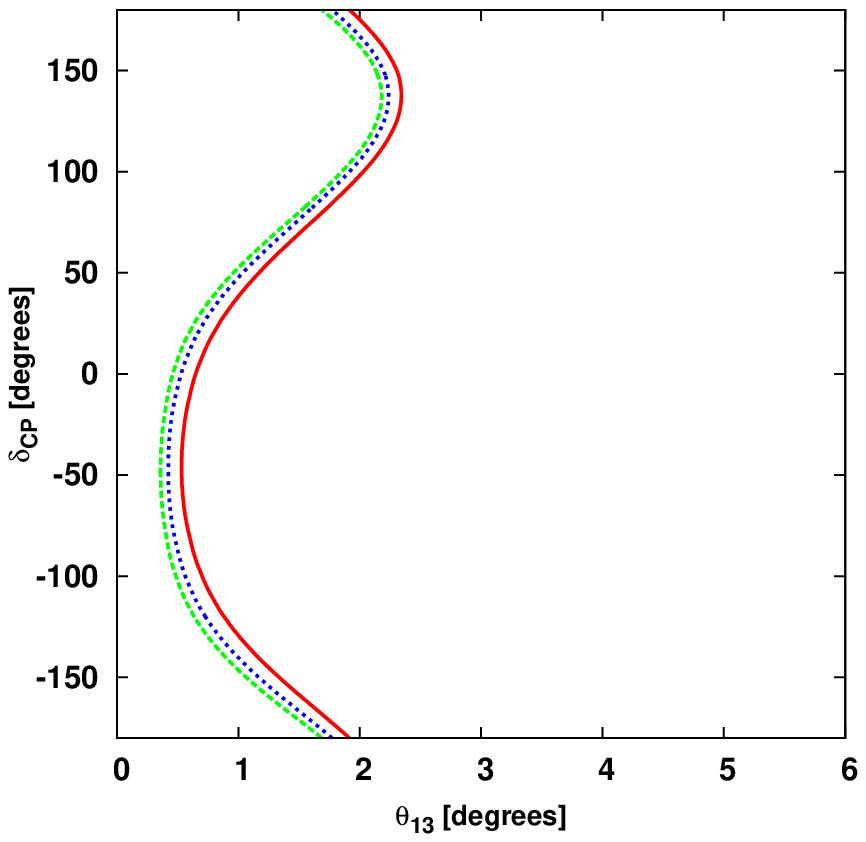}} \hspace{-2cm}
\subfigure{\includegraphics[width=9.5cm]{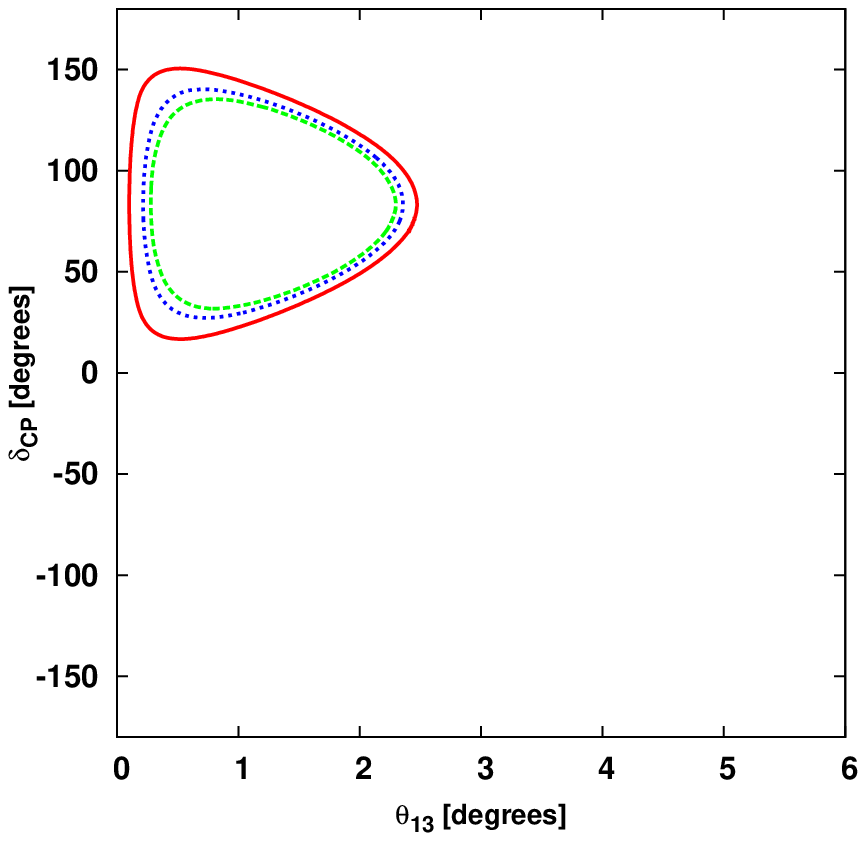}}
\end{center}
\vspace{-5mm}
\caption{90\%, 95\% and 99\% CL contours for setup I (left panels) and
  setup II (right panels). The parameters $\theta_{13}=1^{\circ}$ and
  $\delta=90^{\circ}$ have been taken assuming normal mass ordering and
  $\theta_{23}=45^{\circ}$. The upper row is the contribution of the
  beta-beam, the middle row is that of the electron capture channel
  with the lower row being the combination.}
\label{Fi:C166}
\end{figure}

\subsection{Sensitivity to $\theta_{13}$ and $\delta$}

{\bf Setups I and II} 

Setups I and II use a low boost factor, $\gamma=166$ and a 50~kton
detector, implying relatively low count rates. We find that the
sensitivity to $\theta_{13}$ and $\delta$ is very limited as a
consequence. Violation of CP can be established only for a small range 
of values of the $\delta$ phase and only if  $\theta_{13}$ is close to
the present bounds. The beta-beam channel contributes very little to
the overall sensitivity of the setup. This is due to the
$\gamma^{2}$-dependence of a beta-beam flux. The small flux, when
combined with the small cross-sections at the energies centred on
second oscillation maximum, supplies a scarce count rate. The bulk of
the sensitivity is therefore due to the electron capture channel
placed on first oscillation maximum, as seen in Fig.~\ref{Fi:C166}. The
performance of the beta-beam (upper row), electron capture (middle row)
and their combination (lower row) is shown for $\theta_{13} =
1^{\circ}$ and $\delta = 90^{\circ}$ for setup I (left column) and 
setup II (right column), at 90\%, 95\% and 99\% confidence level (CL). 
 
\vs

{\bf Setups III and III-WC} 

We have also examined the effect of placing the electron capture
beam in the tail of first oscillation maximum. This gives the
beta-beam coverage of the second oscillation maximum and substantial
portions of the first oscillation maximum. For the CERN-Canfranc
baseline and a boost $\gamma =369$, we find that the roles of electron
capture and the beta-beam are reversed compared to $\gamma=166$. The
beta-beam spectrum is peaked around $\sim 1$~GeV while the EC energy
is of 2.55~GeV. As shown in Fig.~\ref{F:ECprobs}, the neutrino
beta-beam spectrum is a good fit for the appearance probability at 
Canfranc, its peak sitting around the first oscillation maximum. The
beta-beam now contributes much more to the sensitivity as it provides
substantial information from the first oscillation maximum and a much
higher count rate from the second oscillation maximum. 

\begin{figure}
\begin{center}
\hspace{-1.5cm}
\subfigure{\includegraphics[width=9.5cm]{PatSII-1-90-BB.eps}} \hspace{-2cm}
\subfigure{\includegraphics[width=9.5cm]{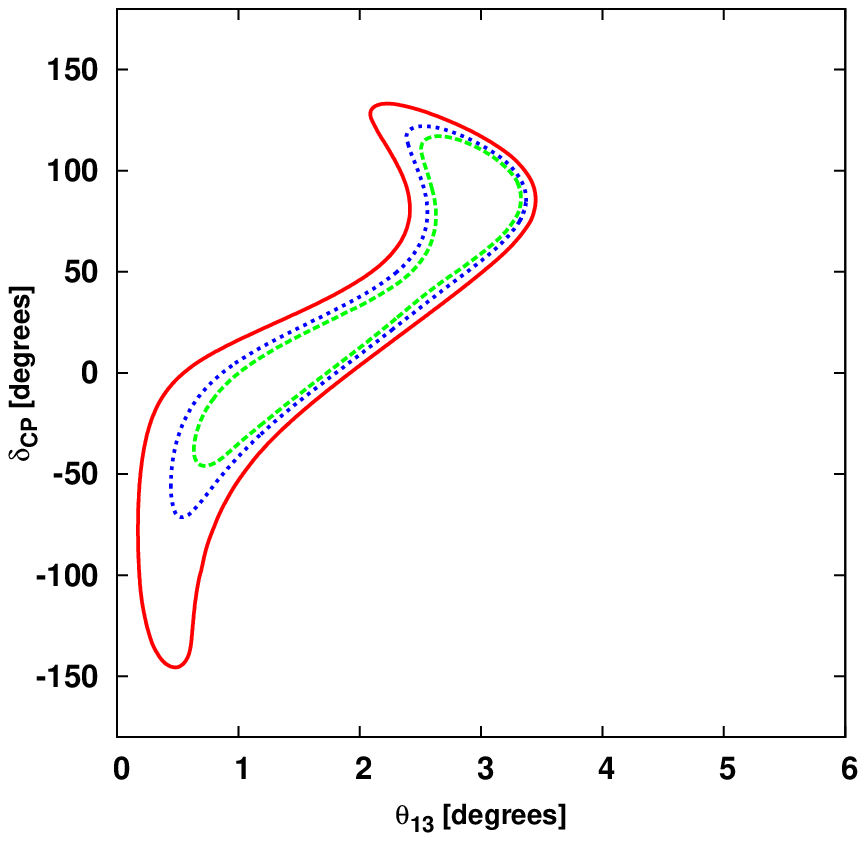}}\\
\hspace{-1.5cm}
\subfigure{\includegraphics[width=9.5cm]{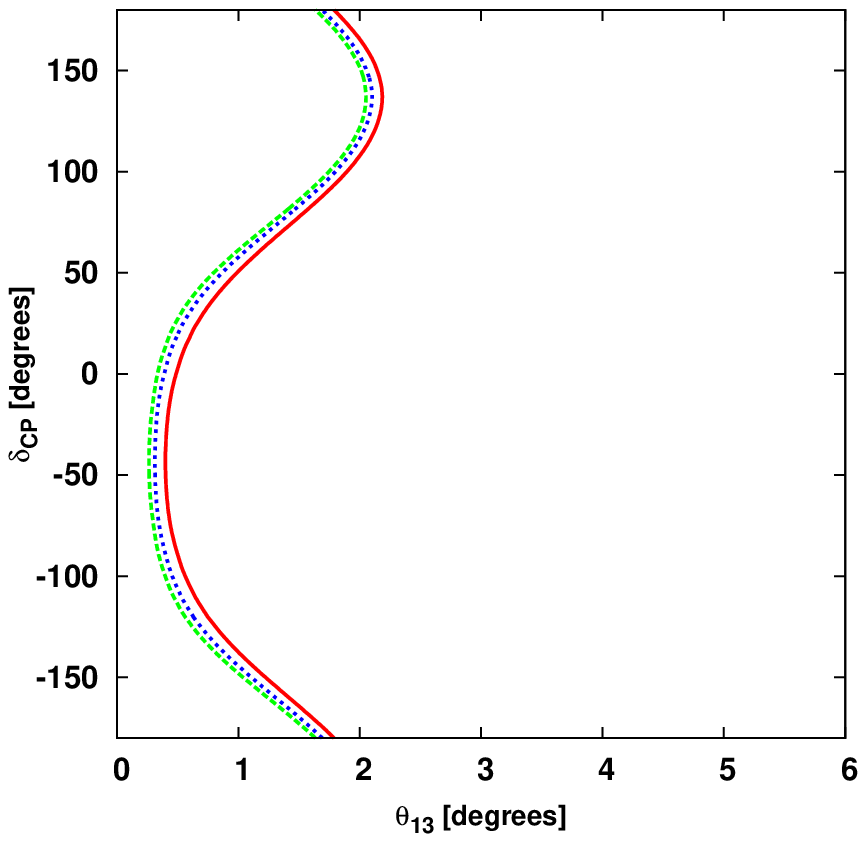}} \hspace{-2cm}
\subfigure{\includegraphics[width=9.5cm]{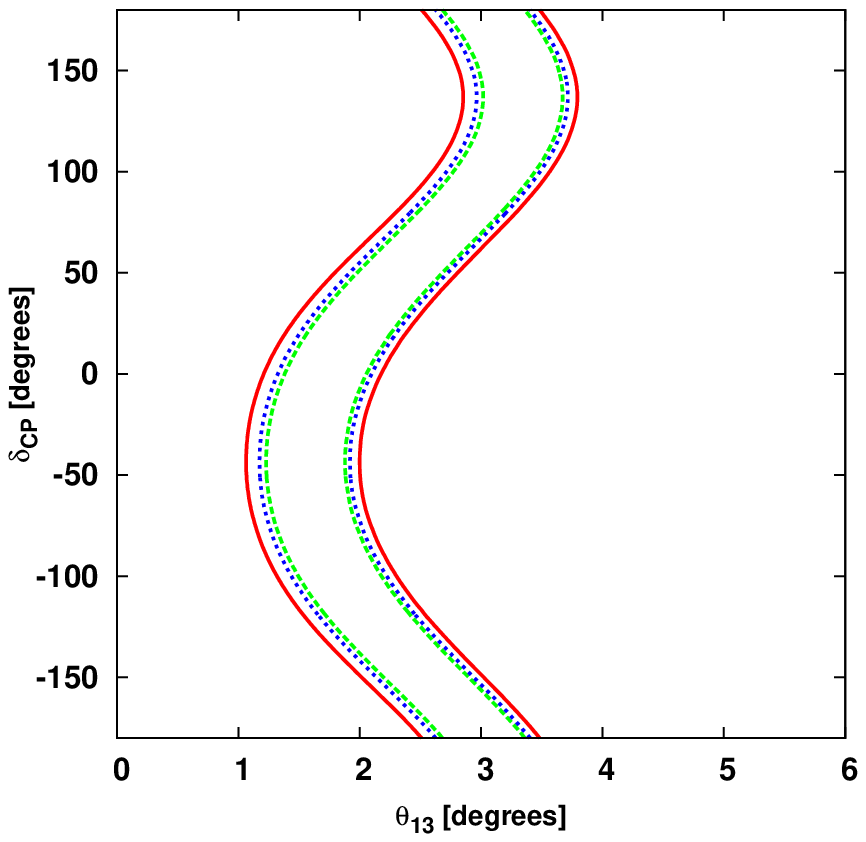}}\\
\hspace{-1.5cm}
\subfigure{\includegraphics[width=9.5cm]{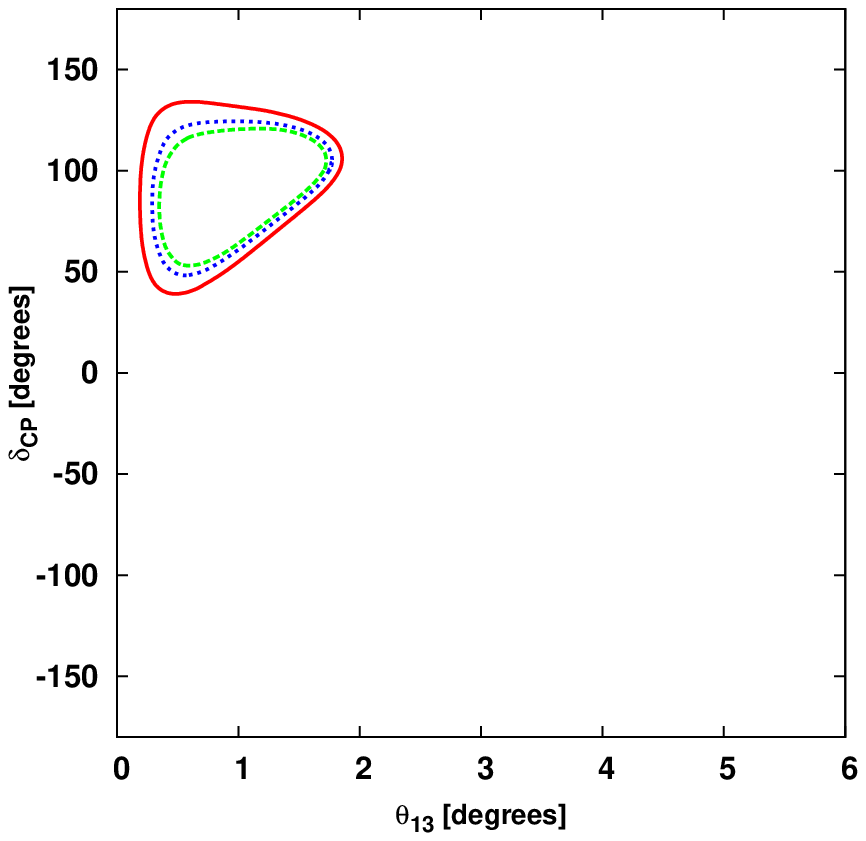}} \hspace{-2cm}
\subfigure{\includegraphics[width=9.5cm]{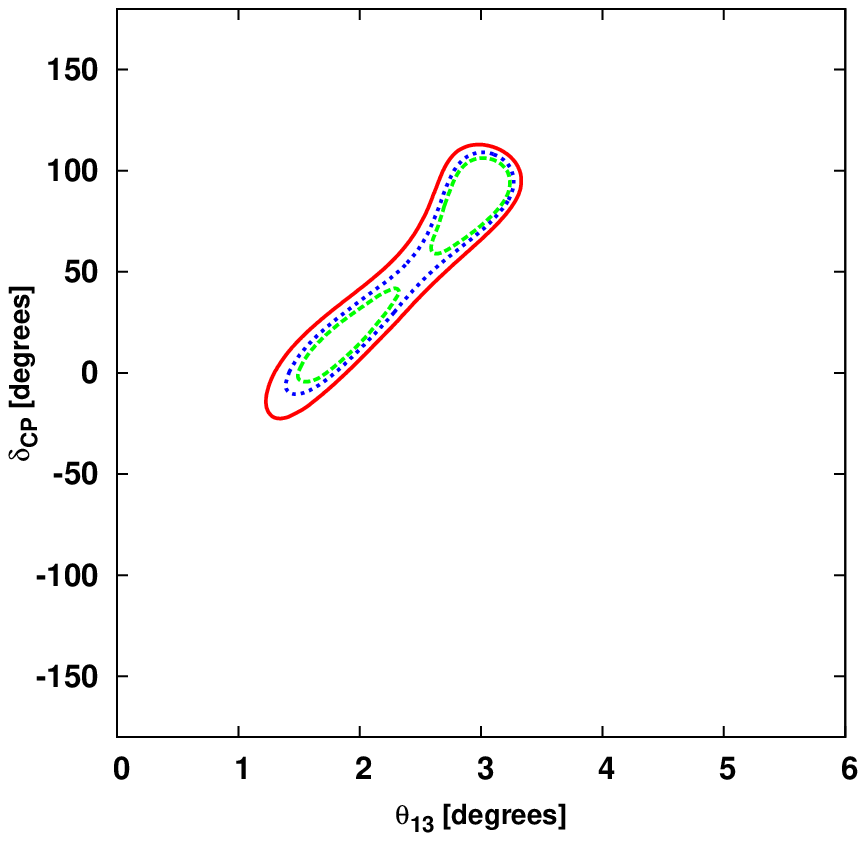}}
\end{center}
\vspace{-5mm}
\caption{90\%, 95\% and 99\% CL contours for the setup III. The left
  column is simulated for $\theta_{13}=1^{\circ}$ and
  $\delta=90^{\circ}$ assuming normal mass ordering and
  $\theta_{23}=45^{\circ}$. The right column is the same but for
  $\theta_{13}=3^{\circ}$. The upper row is the contribution of the
  beta-beam, the middle row is the electron capture channel with the
  lower row being the combination.}
\label{Fi:CC369} 
\end{figure}

\begin{figure}
\begin{center}
\hspace{-1.5cm}
\subfigure{\includegraphics[width=9.5cm]{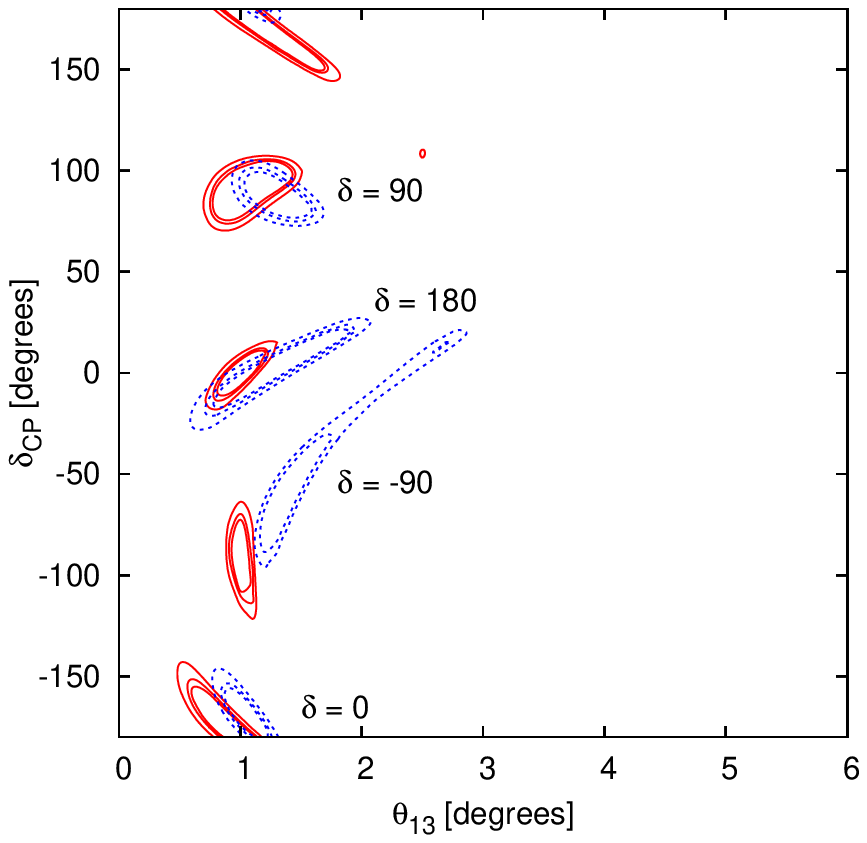}}
\hspace{-2cm}
\subfigure{\includegraphics[width=9.5cm]{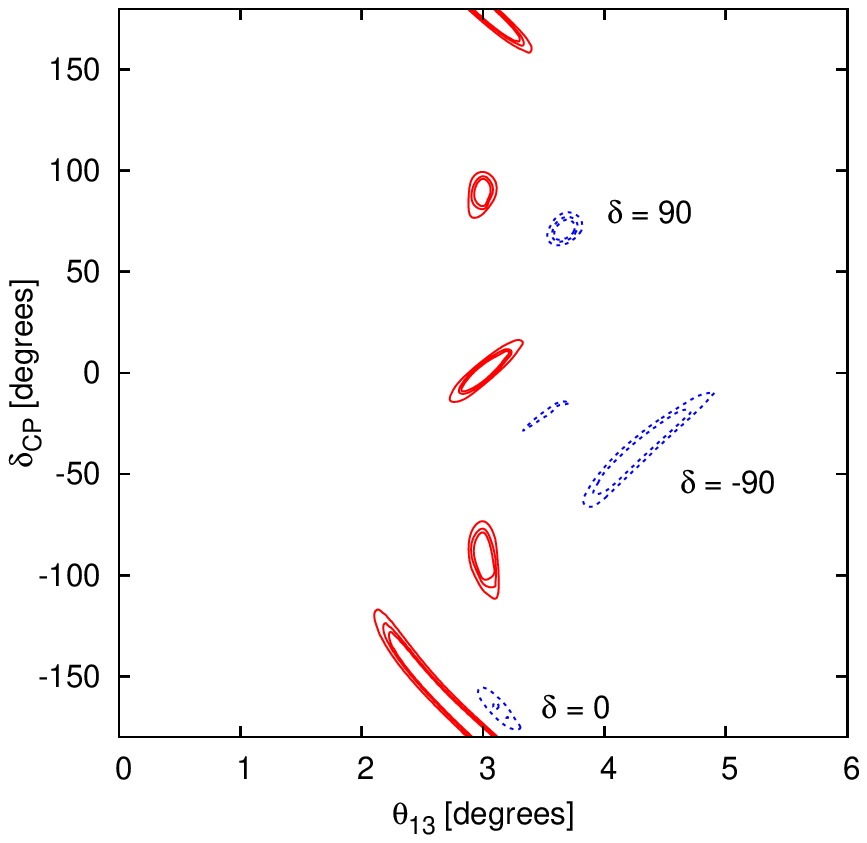}}
\end{center}
\caption{90\%, 95\% and 99\% CL contours for setup III-WC with
  solutions from discrete degeneracies included for
  $\theta_{13}=1^{\circ}$ (left panel) and $\theta_{13}=3^{\circ}$
  (right panel) for different values of the CP-phase, $\delta =
  -90^{\circ}, 0^{\circ}, 90^{\circ}, 180^{\circ}$.}
\label{Fi:CC369clone} 
\end{figure}

In Fig.~\ref{Fi:CC369} and Fig.~\ref{Fi:CC369clone} we show the 90\%,
95\% and 99\% CL contours for the setup III and setup III-WC,
respectively, assuming that the hierarchy is normal and $\theta_{23} =
45^{\circ}$. In Fig.~\ref{Fi:CC369}, we present contours for 
$\delta=90^{\circ}$ and for the cases $\theta_{13}=1^{\circ}$ (left
column) and $3^{\circ}$ (right column). Similarly to
Fig.~\ref{Fi:C166}, in Fig.~\ref{Fi:CC369}, we have included the
contributions from both the $\beta^{+}$-decay (upper row) and electron
capture (middle row) channels separately to investigate their relative
impact, in addition to the total sensitivity (lower row). 

Fig.~\ref{Fi:CC369clone} shows the results for $\theta_{13}=1^{\circ}$
(left column) and $3^{\circ}$ (right column) and four values for
$\delta$, including the hierarchy clone solution. Comparing the
results for the setup III and setup III-WC, one can understand the
effect of the event rate. We see that the 0.5~Mton WC detector gives 
a much better resolution, although not as much as one would
na\"{\i}vely expect from its larger size. As commented above, the
substantially bigger size of the detector cannot be fully exploited
due to the fast drop of the QE cross section at energies above 1~GeV,
where the first oscillation maximum lies.

The full power of the combination between the beta-beam spectrum and
the EC channel is best illustrated in Fig.~\ref{Fi:CC369}. We see that
each of the two techniques separately suffer from a continuum of
solutions. The shape of the allowed region in the $\theta_{13}$ and 
$\delta$ plane can be easily understood by looking at the form of the
oscillation probability in Eq.~(\ref{eq:probappr}). Owing to the
relatively short baseline and small matter effects, we can neglect for
simplicity the matter potential term, $A$, in the following. As we
measure the probability in only one polarity, for fixed energy and
baseline, as is the case for the EC signal, we can relate the allowed
values of $\delta$ and $\theta_{13}$ from a measurement of the
probability as
\begin{equation}
\sin 2 \theta_{13} \simeq -
\frac{\Delta_{21}}{\Delta_{31}}\frac{\Delta_{31} L}{\sin
  (\Delta_{31} L/2)} \cos \Big( \frac{\Delta_{13} L}{2} + \delta \Big)
+ k ~,
\label{sindeg}
\end{equation}
where $k$ is a constant which depends on the true values of
$\theta_{13}$ and $\delta$. This approximation is valid as far as 
$\sin^2 2 \theta_{13} \gg \mbox{few} \ \times 10^{-3}$. The form of
the expression in Eq.~(\ref{sindeg}) matches the continuum of
solutions in Fig.~\ref{Fi:CC369}. In particular, we note
the presence of a minimum for $\theta_{13}$ at $\delta = - \Delta_{13}
L/2$. This result is more general then the approximated form of $\sin
\theta_{13}$ in Eq.~(\ref{sindeg}) and holds also for small values of
$\sin \theta_{13}$. For the choice of the parameters used in
Fig.~\ref{Fi:CC369}, we have $\delta \simeq -44^{\circ}$. The range of
the allowed solutions for $\theta_{13}$ can be computed by looking at
the amplitude of the function in Eq.~\ref{sindeg} and is found to be 
$\Delta \theta_{13} \sim \frac{1}{2} \frac{\Delta_{21}}{\Delta_{31}}
\frac{\Delta_{31} L/2}{\sin (\Delta_{31} L/2)} $. For the choice of
true values in Fig.~\ref{Fi:CC369}, $\Delta \theta_{13} \sim
2^{\circ}$. 

Now consider the contribution of the beta-beam. For simplicity, we
analyse its impact by looking at the energy of the first oscillation
maximum, again neglecting matter effects. As for the previous case,
the minimum  of the continuum solutions for $\theta_{13}$ is located  
at $\delta = - \Delta_{13} L/2=-\pi/2$; shown clearly in
Fig.~\ref{Fi:CC369}.

The power of the combination of the beta-beam and electron-capture
channels is in the difference in phase and in amplitude between
the two fake sinusoidal solutions; their combination selects a
narrow allowed region in the parameter space, much more constrained
then the two separate techniques. This effect is clearly visible in
the lower plots in Fig.~\ref{Fi:CC369}.

For $\theta_{13}=3^{\circ}$, there is still some intrinsic degeneracy
that cannot be completely removed at 99\% CL. The marked difference
between the beta-beam alone and the combination with the electron
capture in this case demonstrates the importance of data from the high
energies. For $\sin^2 2\theta_{13} \sim 10^{-3}$ and $\delta\sim
\pi/2$, the beta-beam configuration is able to determine the allowed
region in the ($\theta_{13}$-$\delta$) parameter space with relatively
good accuracy. For this range of $\theta_{13}$, the dominant
interference term helps in resolving any degeneracy. This is not the
case for other values of $\delta$. 

In Fig.~\ref{Fi:CC369clone}, we show the results for setup III-WC,
with the effects of the hierarchy clone solution taken into
account. From a comparison of Figs.~\ref{Fi:CC369}
and~\ref{Fi:CC369clone}, the increase in event rates improves the
results substantially.  However, owing to the relatively short
distance, $L=650$~km, the mass ordering can be determined only for
large values of the mixing angle $\theta_{13}$ (see below). The
hierarchy degeneracy worsens the ability to measure $\theta_{13}$ and
$\delta$ with good precision, especially for negative true values of
$\delta$. 

\vs

{\bf Setups IV and IV-WC} 

\begin{figure}
\begin{center}
\hspace{-1.5cm}
\subfigure{\includegraphics[width=9.5cm]{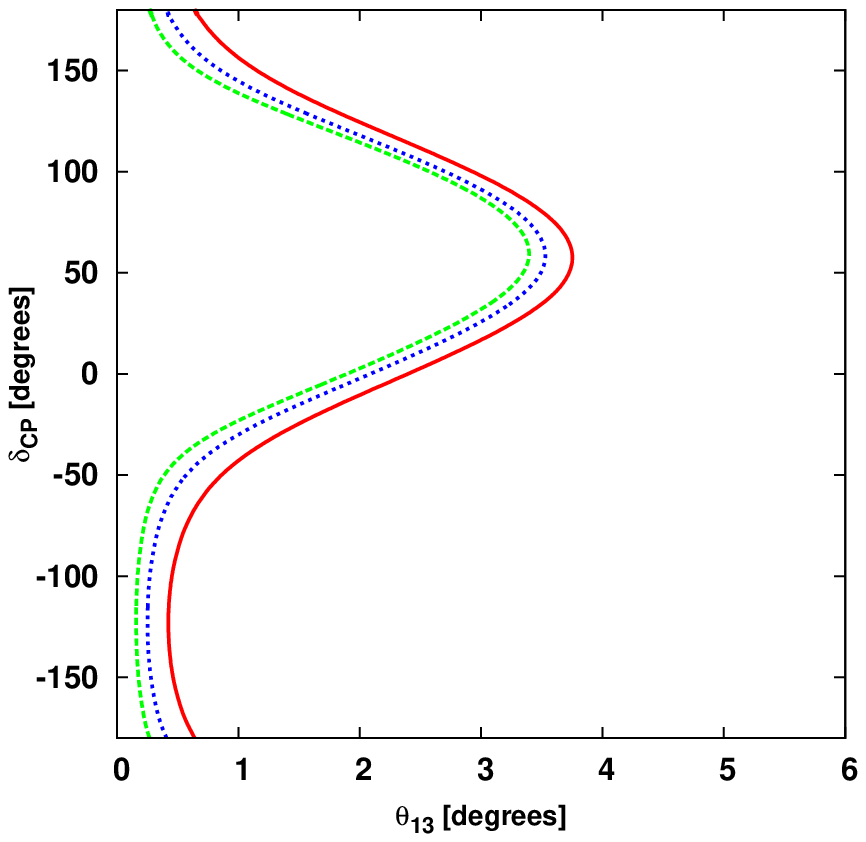}} \hspace{-2cm}
\subfigure{\includegraphics[width=9.5cm]{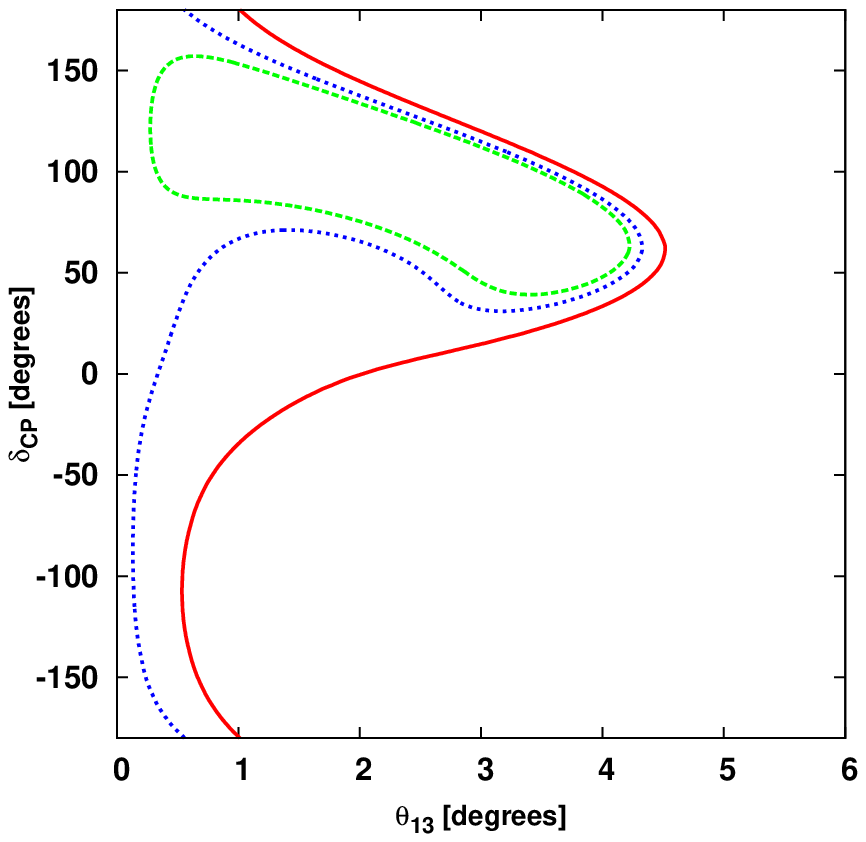}}\\
\hspace{-1.5cm}
\subfigure{\includegraphics[width=9.5cm]{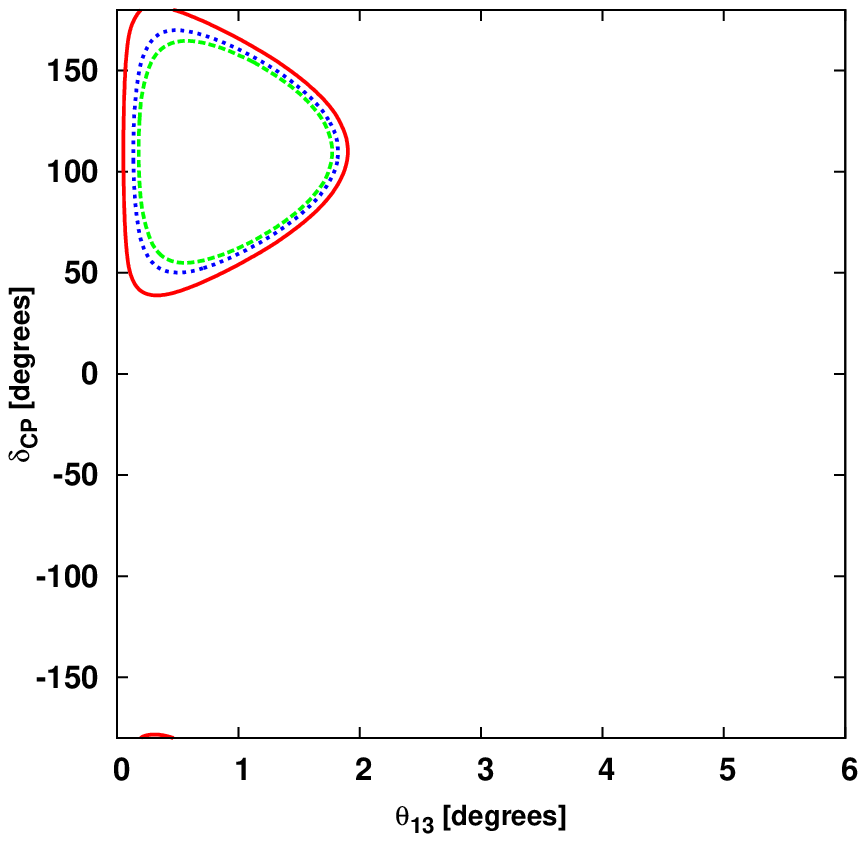}} \hspace{-2cm}
\subfigure{\includegraphics[width=9.5cm]{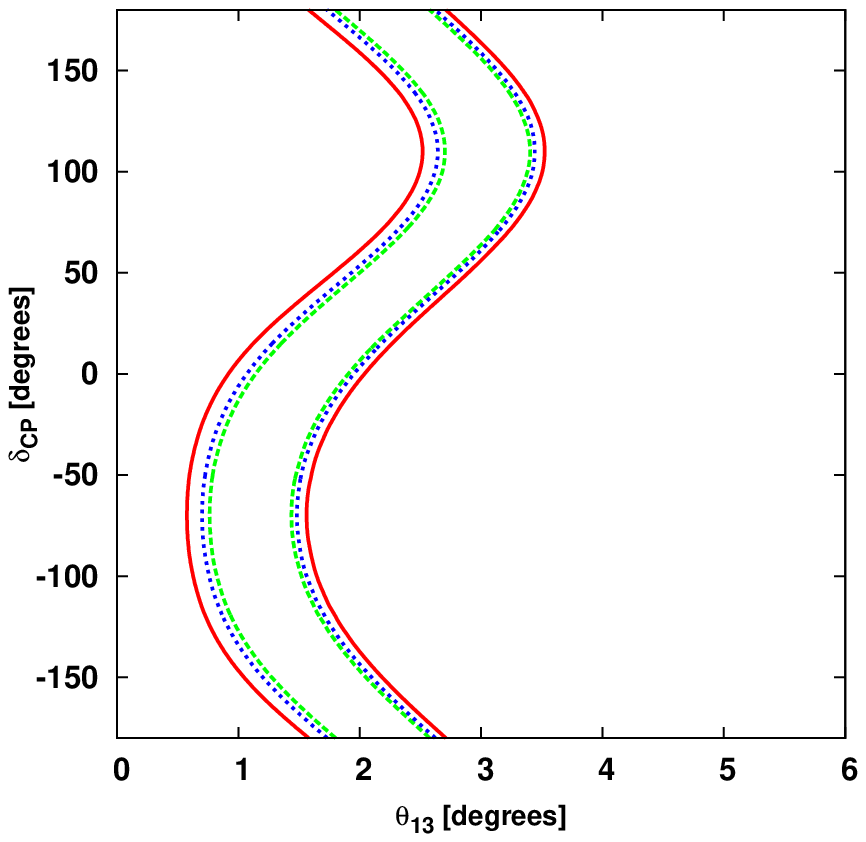}}\\
\hspace{-1.5cm}
\subfigure{\includegraphics[width=9.5cm]{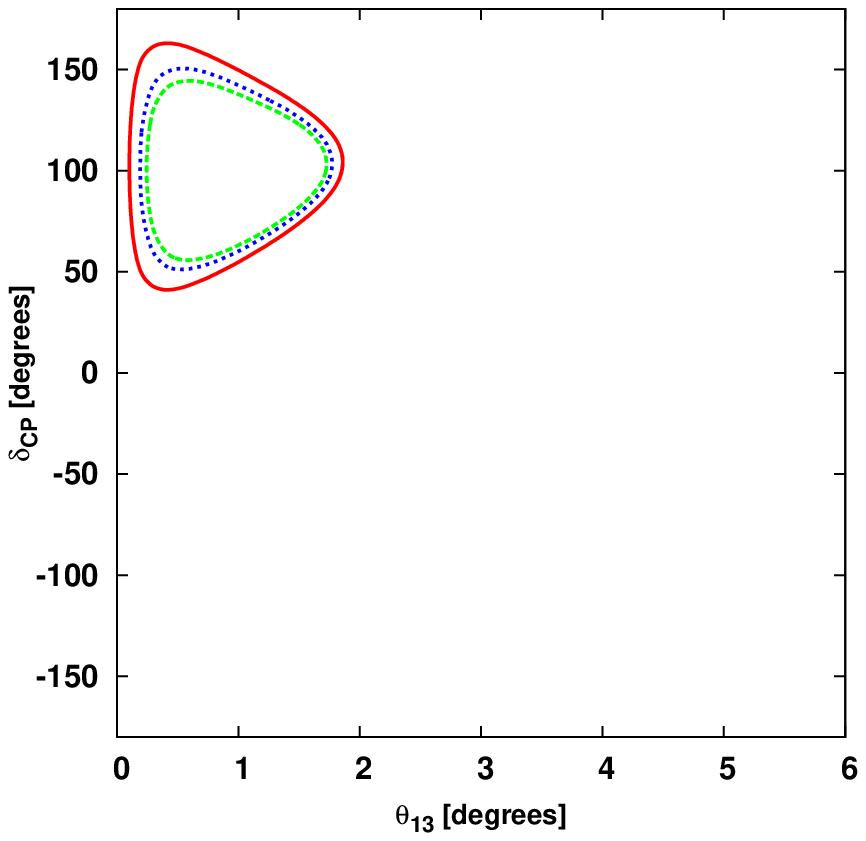}} \hspace{-2cm}
\subfigure{\includegraphics[width=9.9cm]{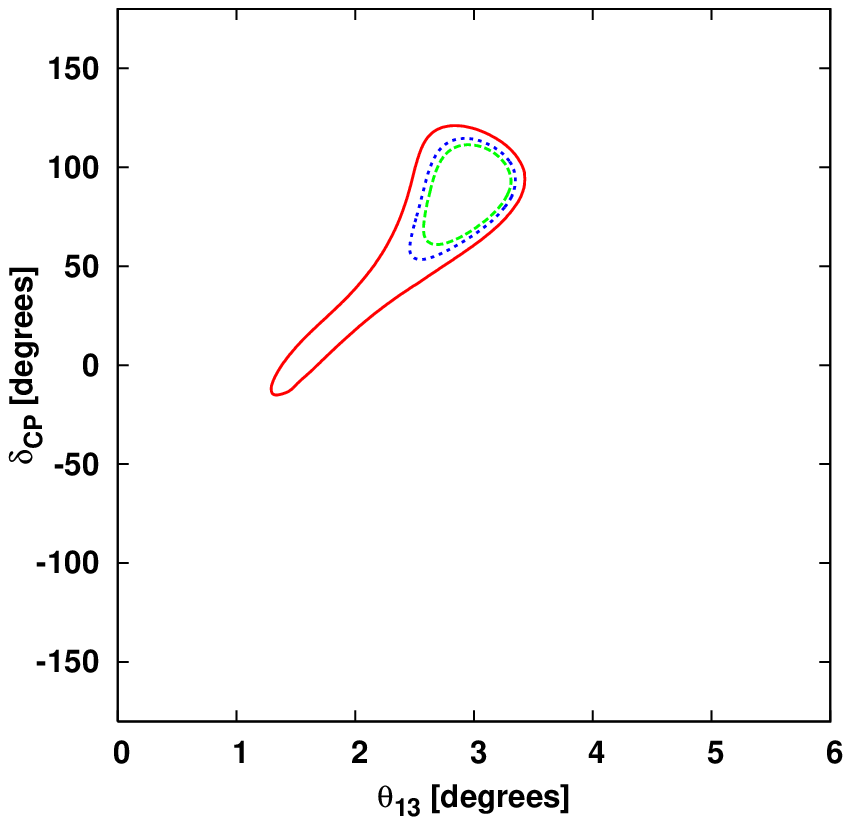}}
\end{center}
\caption{Same as Fig.~\ref{Fi:CC369} but for setup IV.}  
\label{Fi:CB369}  
\end{figure}

The use of Boulby instead of Canfranc for a boost of $\gamma=369$
serves as an intermediate case with respect to the last two setups in
the sense that the position of the electron capture is neither on
first oscillation maximum or far into the tail. Consequently, the  
beta-beam has some, but not a considerable coverage of the first
oscillation region. Boulby provides a much longer baseline,
$L=1050$~km. This has two contrasting effects on the sensitivity to 
measure CP-violation: on one side it provides sufficient matter effects
to resolve the hierarchy degeneracy even for small values of
$\theta_{13}$; on the other, it decreases the available statistics
with respect to Canfranc.

\begin{figure}
\begin{center}
\hspace{-1.5cm}
\subfigure{\includegraphics[width=9.5cm]{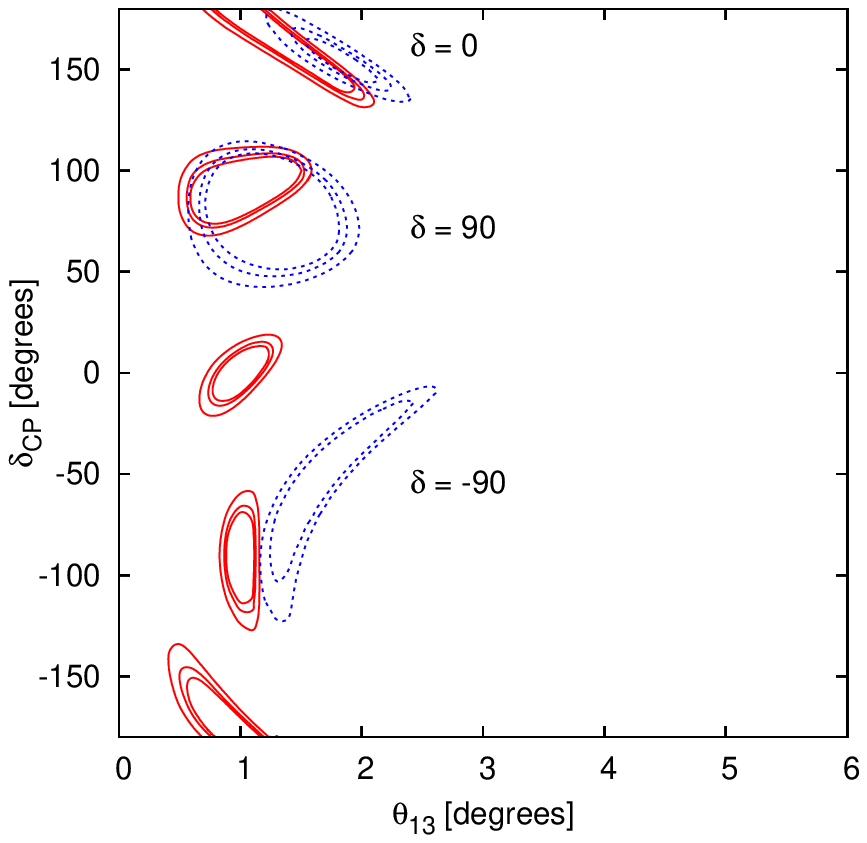}}
\hspace{-2cm}
\subfigure{\includegraphics[width=9.5cm]{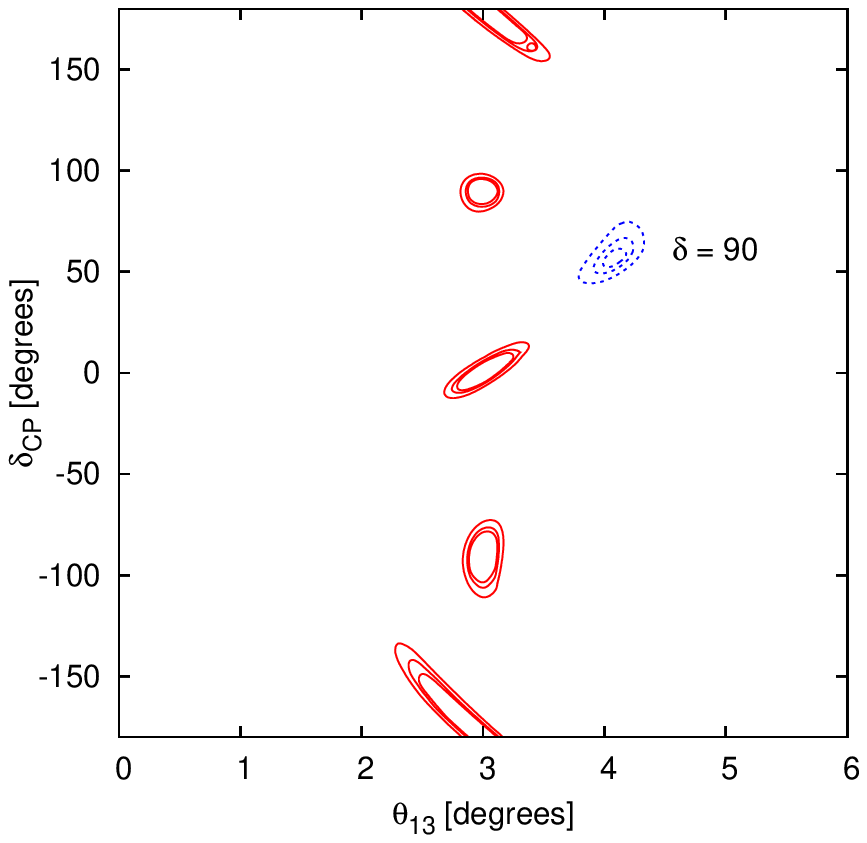}}
\end{center}
\caption{Same as Fig.~\ref{Fi:CC369clone} but for setup IV-WC.}
\label{Fi:CB369clone}
\end{figure}

In Fig.~\ref{Fi:CB369} and Fig.~\ref{Fi:CB369clone}, we report the
results for setup IV and setup IV-WC, respectively, for normal mass
hierarchy and for the cases $\theta_{13} = 1^{\circ}$ (left panels) and
$\theta_{13}=3^{\circ}$ (right panels). In Fig.~\ref{Fi:CB369}, we
show contours for $\delta=90^{\circ}$. Similarly to Figs.~\ref{Fi:C166}
and~\ref{Fi:CC369}, in Fig.~\ref{Fi:CB369} we present the results for
the $\beta^{+}$-decay flux only (upper row), electron capture flux
only (middle row) and their combination (lower
row). Fig.~\ref{Fi:CB369clone} shows the results for four values for 
$\delta$, including the hierarchy clone solution (blue dashed
contours). The electron capture channel displays similar behaviour to 
the Canfranc high boost case (setup III, middle row of
Fig.~\ref{Fi:CC369}) at $\theta_{13}=3^{\circ}$. The minimum value of
the continuum solutions of $\theta_{13}$ is located at $\delta = -
66^{\circ}$ and the allowed range of values is $\sim 2^{\circ}$. For
the beta-beam (upper rows), the first oscillation maximum energy bin
contributes only marginally to the overall sensitivity, as can be
understood from Fig.~\ref{F:ECprobs}. Therefore, the allowed region in
Fig.~\ref{Fi:CB369} has a different shape with respect to the case
shown in Fig.~\ref{Fi:CC369}. In addition, the smaller count rate
results in a poorer resolution. However, the synergy between beta-beam
and electron capture is more important here. The lack of concurrence
of the beta-beam and electron capture allowed regions implies that
their combination constrains $\delta$ and $\theta_{13}$ in small
ranges. This can be clearly seen in Fig.~\ref{Fi:CB369}.

In addition, the longer baseline allows for a good determination of
the mass ordering (see below), eliminating more degenerate solutions
and providing an improved sensitivity to CP-violation with respect to
setups III and III-WC. Comparing Fig.~\ref{Fi:CB369clone} with the 
lower row of Fig.~\ref{Fi:CB369}, the improvement from setup
IV to setup IV-WC is noticeable.

\subsection{CP-violation discovery potential}

We now consider the CP-discovery potential for the various setups. For
the low-$\gamma$ options, setups I and II, the sensitivity is very
limited, in agreement with the findings already reported in
Fig.~\ref{Fi:C166}. Henceforth, we will not show results of these two
setup as they always possess worse physics reach compared to the other
setups.

\begin{figure}
\begin{center}
\subfigure{\includegraphics[width=8cm]{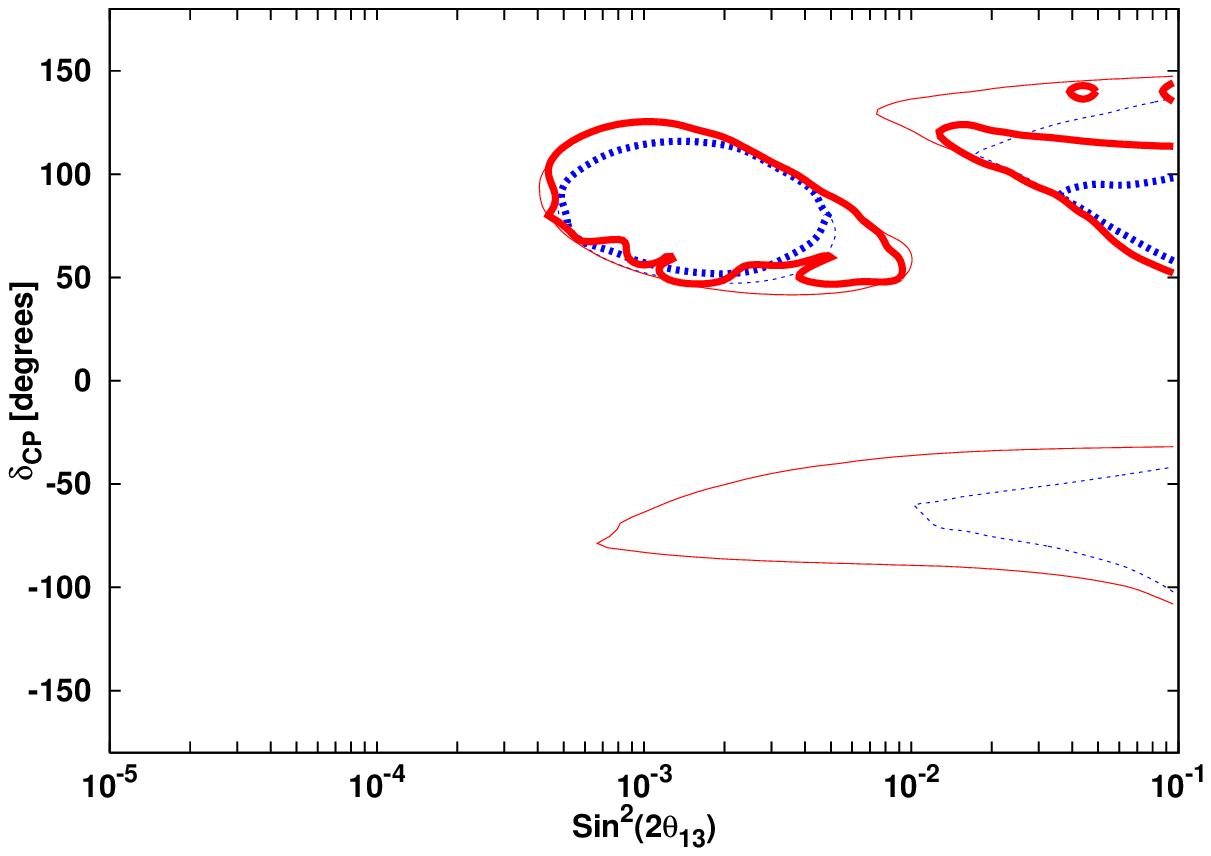}} 
\subfigure{\includegraphics[width=8cm]{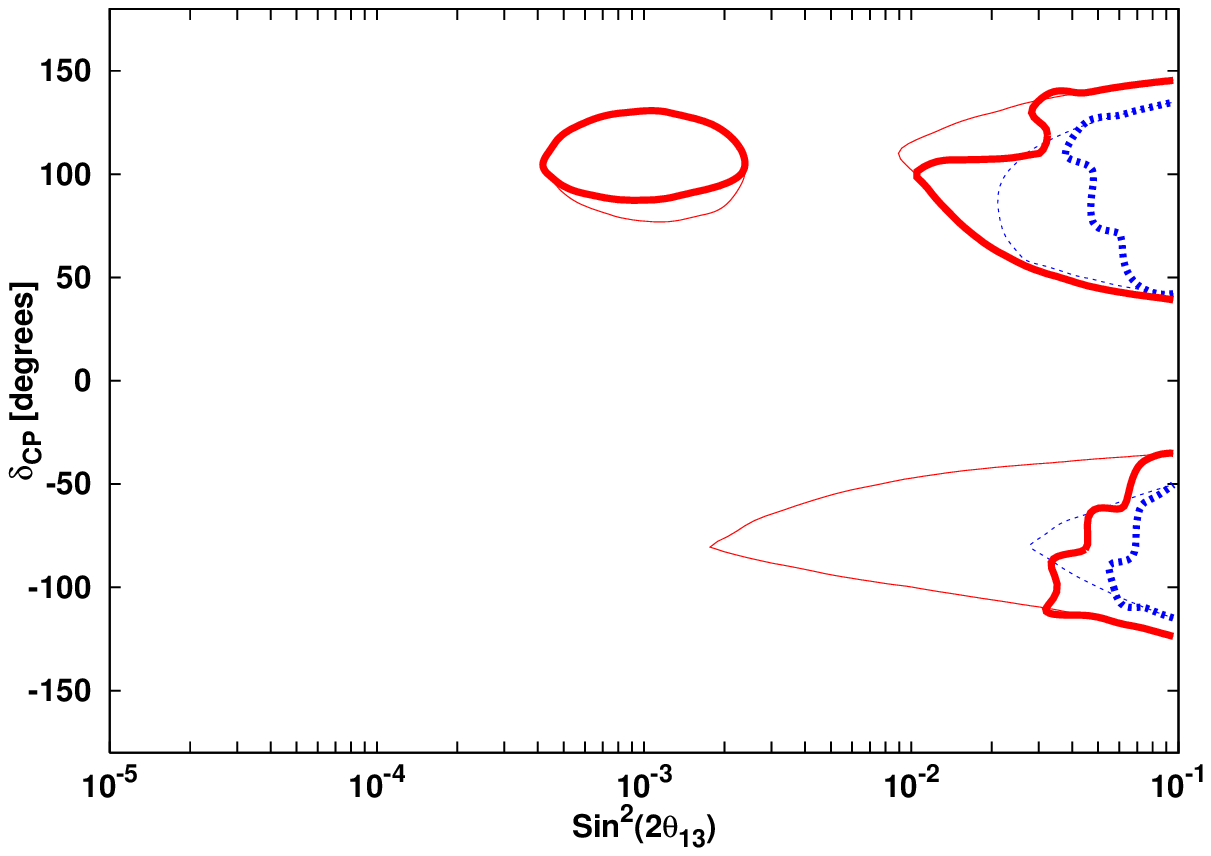}}\\ 
\end{center}
\caption{CP-violation discovery potential at 99\% CL for setup III
  (left panel) and IV (right panel). In each case, we present the
  results for the beta-beam only (blue dotted lines) and the
  combination with the electron capture result (red solid lines), both 
  without (thin lines) and with (thick lines) taking the hierarchy
  degeneracy into account.}
\label{Fi:CP50kt}
\end{figure}

\begin{figure}
\begin{center}
\subfigure{\includegraphics[width=8cm]{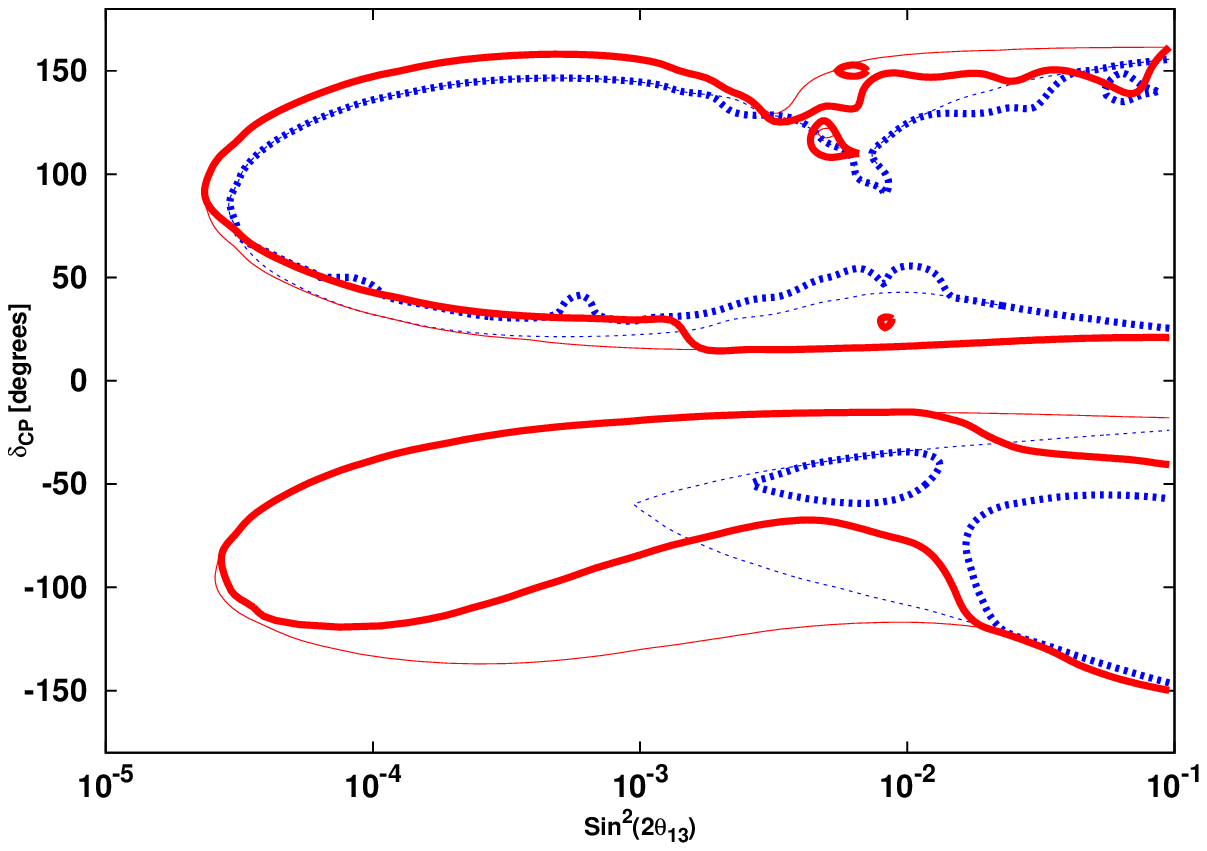}} 
\subfigure{\includegraphics[width=8cm]{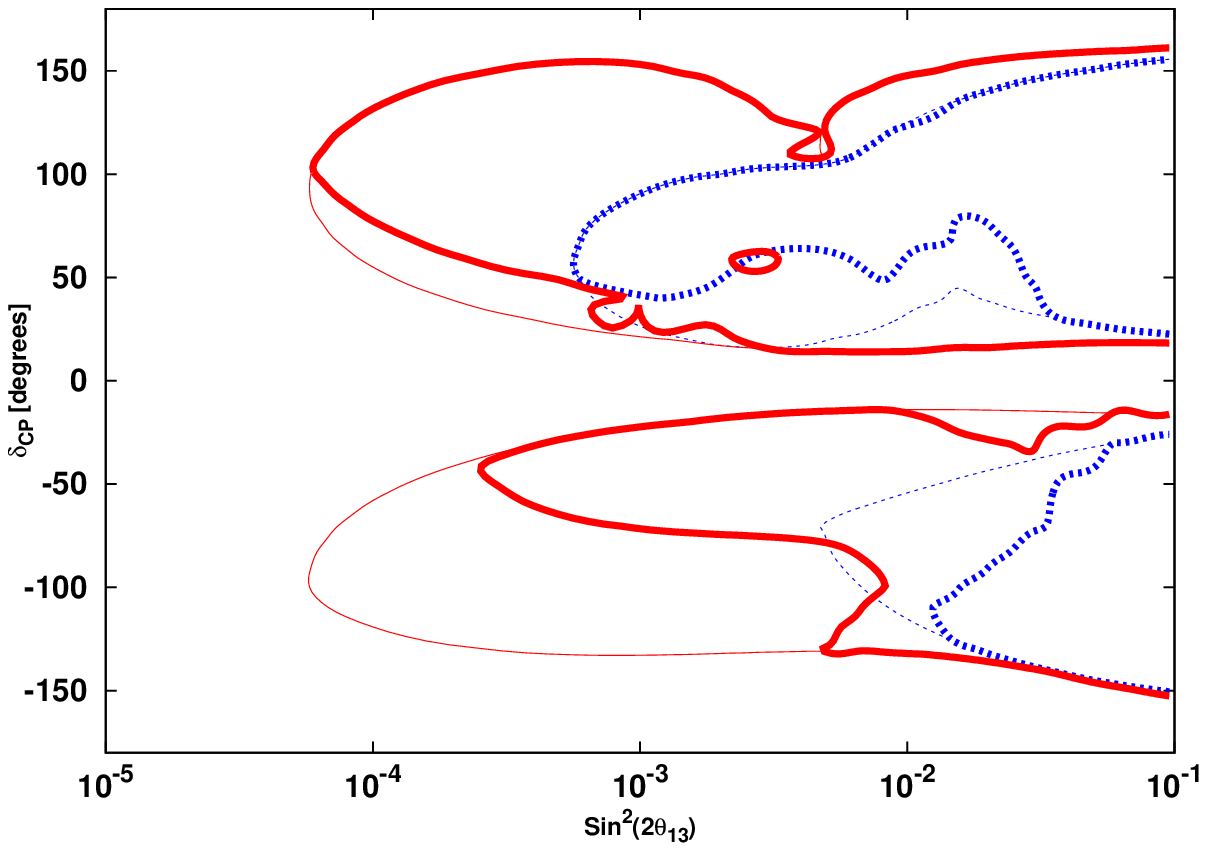}}\\ 
\end{center}
\caption{Same as Fig.~\ref{Fi:CP50kt} but for setup III-WC (left
  panel) and setup IV-WC (right panel).}
\label{Fi:CPWC}
\end{figure}

Setups III, III-WC, IV and IV-WC have a much better physics reach, as
shown in Fig.~\ref{Fi:CP50kt} and Fig.~\ref{Fi:CPWC}, where the
CP-violation discovery potential at 99\% CL for the 50 kton TASD or
LAr detector (setups III and IV) and 0.5~Mton WC detector (setups
III-WC and IV-WC) are depicted, respectively. In both figures, the 
CERN-Canfranc baseline is displayed in the left panel and the
CERN-Boulby baseline in the right panel. In each panel, we present the 
results for the beta-beam only (blue dotted lines) and the combination
with the electron capture result (red solid lines), both  without
(thin lines) and with (thick lines) taking the hierarchy degeneracy
into account. In all cases we note that the addition of the EC channel
weakens the impact of the intrinsic degeneracy, significantly
improving the sensitivity. The CP-discovery potential depends on
various factors, mainly the available count rate and the presence of
the hierarchy clone degeneracy. The count rates are important, as can
be understood by comparing the results for the setups III with III-WC
and IV with IV-WC. However, it should be pointed out that the WC
detector is not optimised for the high energies, where the QE cross
section is small and the information on the energy for the total cross
section is poor. We could expect a similar sensitivity to CP-violation
for a TASD or LAr detector with exposure a factor only a few times
larger than the one considered in this analysis.

The effects of the hierarchy degeneracy are important, significantly
more so for $\delta<0$, where there is a loss of sensitivity to
CP-violation by a couple of orders of magnitude in $\sin^2
2\theta_{13}$. We can understand this effect by looking at
Fig.~\ref{Fi:CC369clone}. For the shorter CERN to Canfranc baseline,
we note that for positive $\delta$ either the hierarchy can be
determined or, where not, the hierarchy clone solution significantly 
overlaps with the true one. In the case of negative values of $\delta$,
the hierarchy cannot be resolved even for large values of
$\theta_{13}$; the clone solution stretches into CP-conserving values
preventing the possibility to determine CP-violation. The inclusion of
the high energy EC channel helps in resolving the degeneracy, with
CP-discovery down to $\sin^{2}2\theta_{13} \sim 3\times 10^{-5}$ at
99\% CL (left panel of Fig.~\ref{Fi:CPWC}).

The CERN to Boulby baseline (setups IV and IV-WC) has stronger
degenerate effects, but it also provides a better ability to resolve
them. The results with and without the hierarchy clone solution are
not significantly different since, for the values for which one has
sensitivity to CP-violation, the hierarchy can be resolved. For setup
IV, the reach is limited to large values of $\sin^2 2 \theta_{13}$.
Using the  WC detector, the much larger count rate brings
significantly better results: CP-violation can be established for a
large fraction of $\delta$ values, even for $\sin^2 2 \theta_{13} \sim
\mbox{few} \ \times 10^{-4}$ at 99\% CL.

Comparing the two locations of the detector, we notice that the
shorter baseline (CERN-Canfranc) has a slightly (significantly) better
reach for CP-violation at positive (negative) values of $\delta$ than
the longer baseline (CERN-Boulby). The longer option, however, performs
slightly better at negative $\delta$ if the hierarchy is known to be
normal and significantly better if the ordering is not determined.
This is because the longer baseline can identify the neutrino mass
hierarchy for these values of $\theta_{13}$, therefore resolving this
degeneracy.

\subsection{Mass hierarchy determination}

\begin{figure}
\begin{center}
\subfigure{\includegraphics[width=8cm]{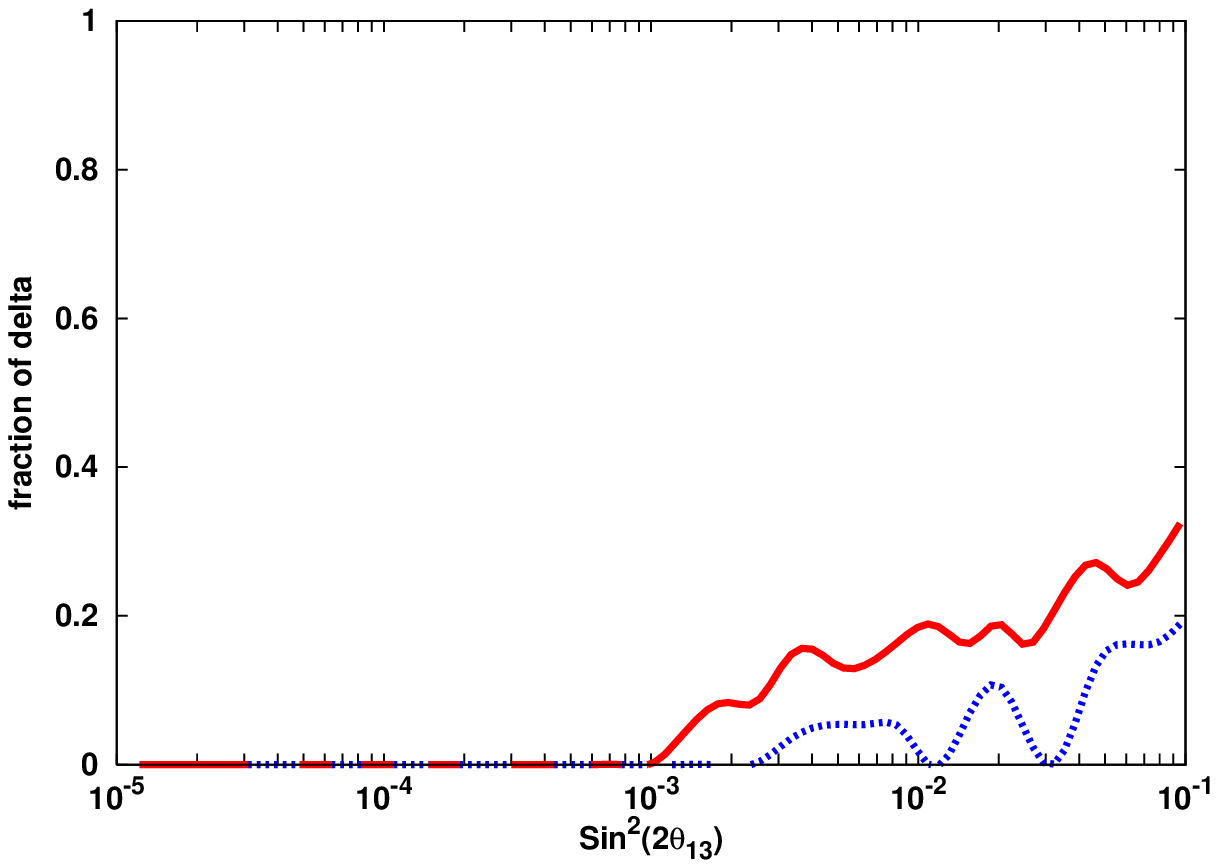}} 
\subfigure{\includegraphics[width=8cm]{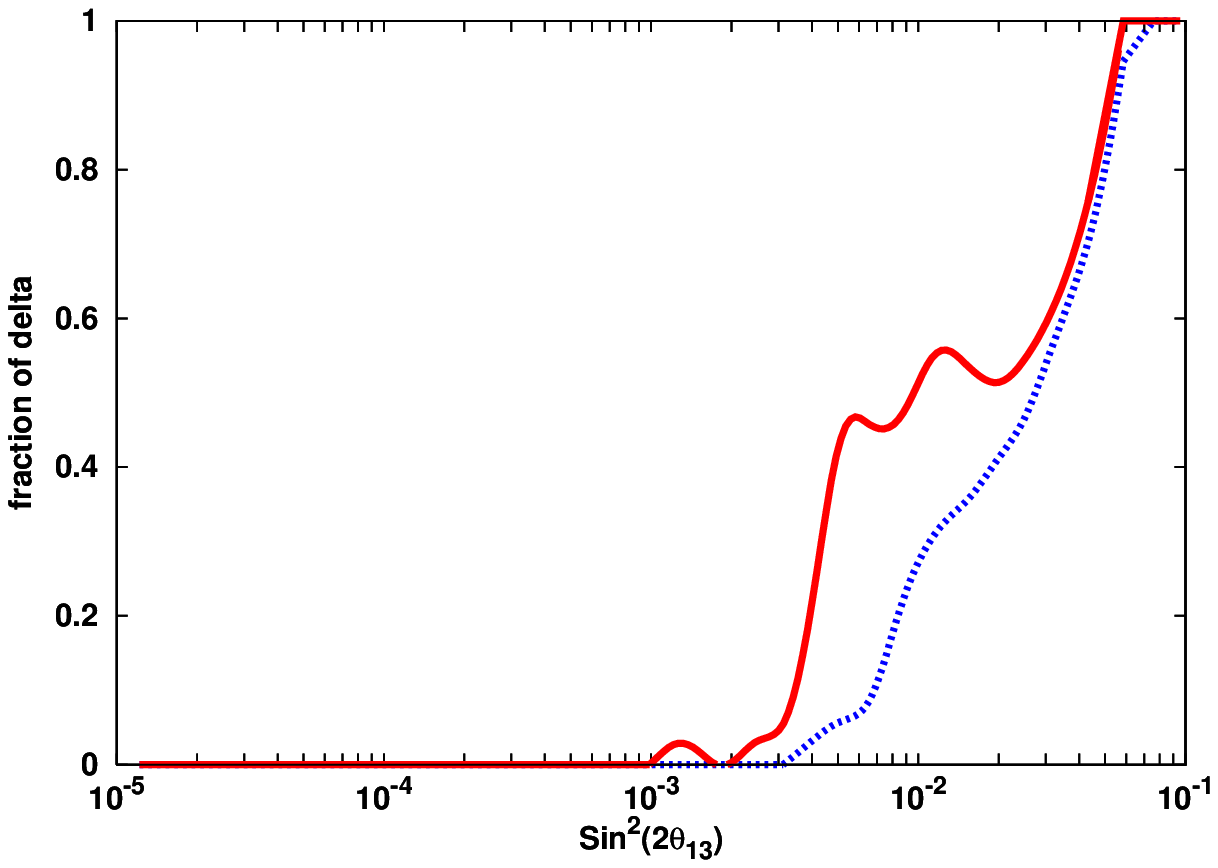}}\\
\end{center}
\caption{Fraction of $\delta$ for which the neutrino mass hierarchy
  can be determined at 99\% CL for setup III-WC (left panel) and IV-WC
  (right panel). In each case, we present the results for the
  beta-beam only (blue dotted lines) and the combination with the
  electron capture result (red solid lines).}
\label{Fi:hierarchyWC}
\end{figure}

In Fig.~\ref{Fi:hierarchyWC}, we present the results for the neutrino
mass hierarchy determination, but only for the setups with a 0.5~Mton
WC detector (setups III-WC and IV-WC). We do not consider the
CERN-Frejus cases; the shorter baseline being unable to distinguish
the type of hierarchy. 

In both cases, the contribution from the beta-beam channel is shown in
blue dashed lines and the result for the combination with the electron
capture channel is shown by the red solid lines. As matter effects are
more important at high energies, we see that the inclusion of the
electron capture flux improves the results, and in particular for the
low values of $\sin^2 2\theta_{13}$ for which the measurement is
possible. However, the chances to determine the mass hierarchy are 
very limited for the CERN-Canfranc baseline, never reaching more than
30\% of the values of the CP-violating phase $\delta$. On the other
hand, the CERN-Boulby baseline, with its larger matter effect,
represents a much more promising setup, for which the determination of
the mass hierarchy would be possible for all values of $\delta$ for
$\sin^2 2\theta_{13} \simeq \mbox{few} \times 10^{-2}$, and with a
50\% probability for  $\sin^2 2\theta_{13} \simeq \mbox{few} \times
10^{-3}$.


\section{Discussion}
\label{sec:discussion}

Let us note that by the time this experiment could possibly take
place, there will be much better knowledge of the neutrino oscillation
parameters, improving quite considerably the results presented
here. Throughout this study, we have adopted a very conservative
stance, namely the assumption that the errors on the neutrino
parameters will remain the same by the time this experiment might
start taking data. However, this is very likely not to be the case.
Here, we present the results when no errors are included on the
assumed values of the neutrino oscillation parameters, i.e., they are
known with perfect precision. The actual performance of the experiment
would lie in between these results and those presented in previous
sections. For brevity, we only consider results for setups III-WC and
IV-WC.

\begin{figure}
\begin{center}
\subfigure{\includegraphics[width=8cm]{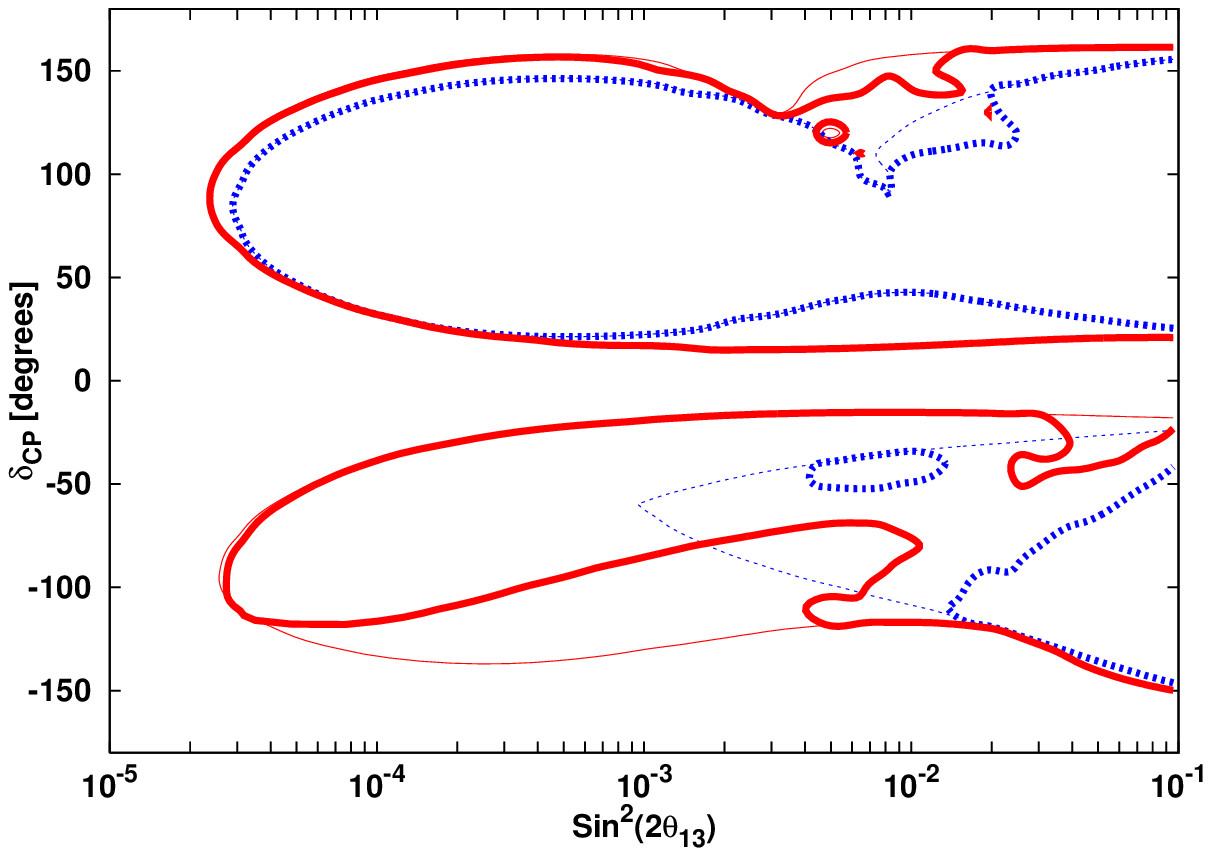}} 
\subfigure{\includegraphics[width=8cm]{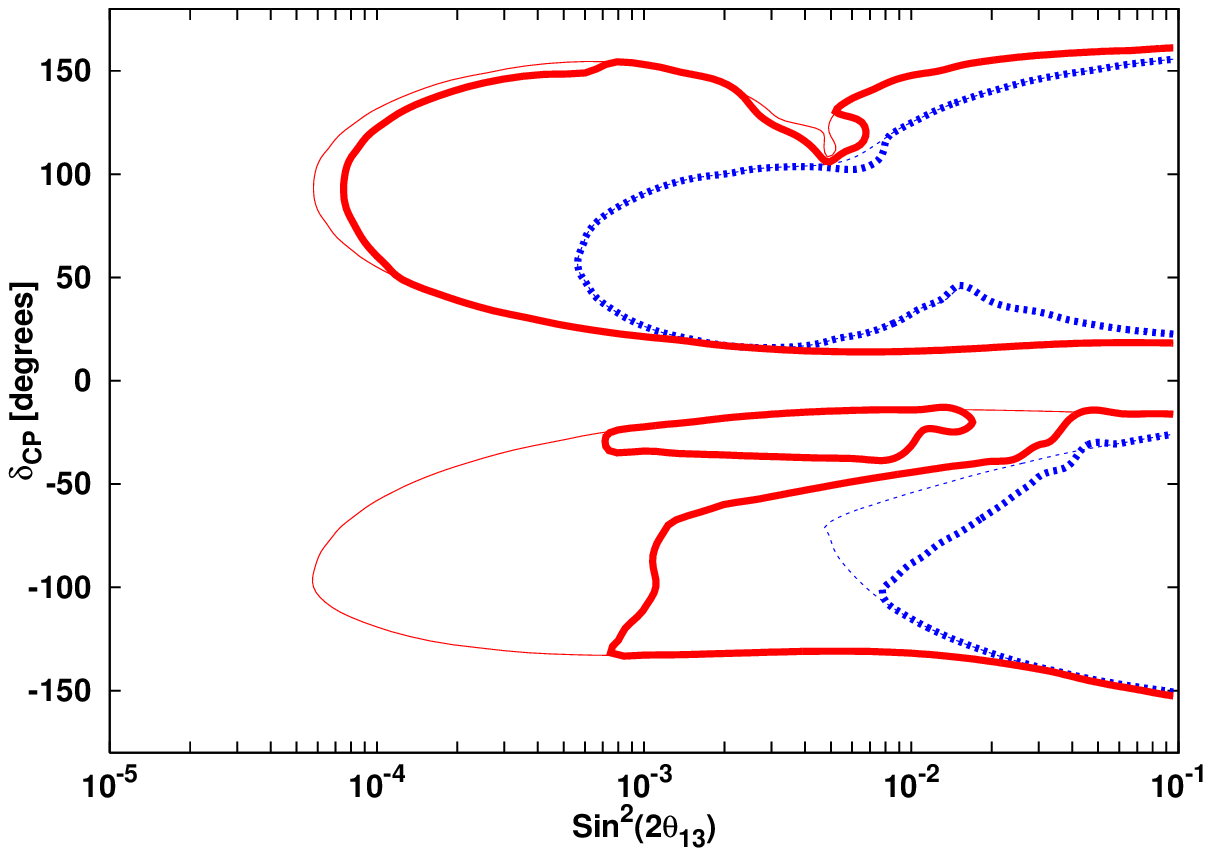}}\\
\end{center}
\caption{Same as Fig.~\ref{Fi:CPWC}, but with negligible error in the
  value of the assumed neutrino parameters.}
\label{Fi:CPnoprior}
\end{figure}

The CP-discovery potential is shown in Fig.~\ref{Fi:CPnoprior}, where
the left (right) panel represents the setup III-WC (setup IV-WC) and
the beta-beam only (beta-beam plus electron capture) performance is
shown by the dashed blue lines (red solid lines). From
Figs.~\ref{Fi:CPWC} and~\ref{Fi:CPnoprior}, the difference between use
of the present uncertainty in the neutrino oscillation parameters and
no uncertainty is seen to be small.

\begin{figure}
\begin{center}
\subfigure{\includegraphics[width=8cm]{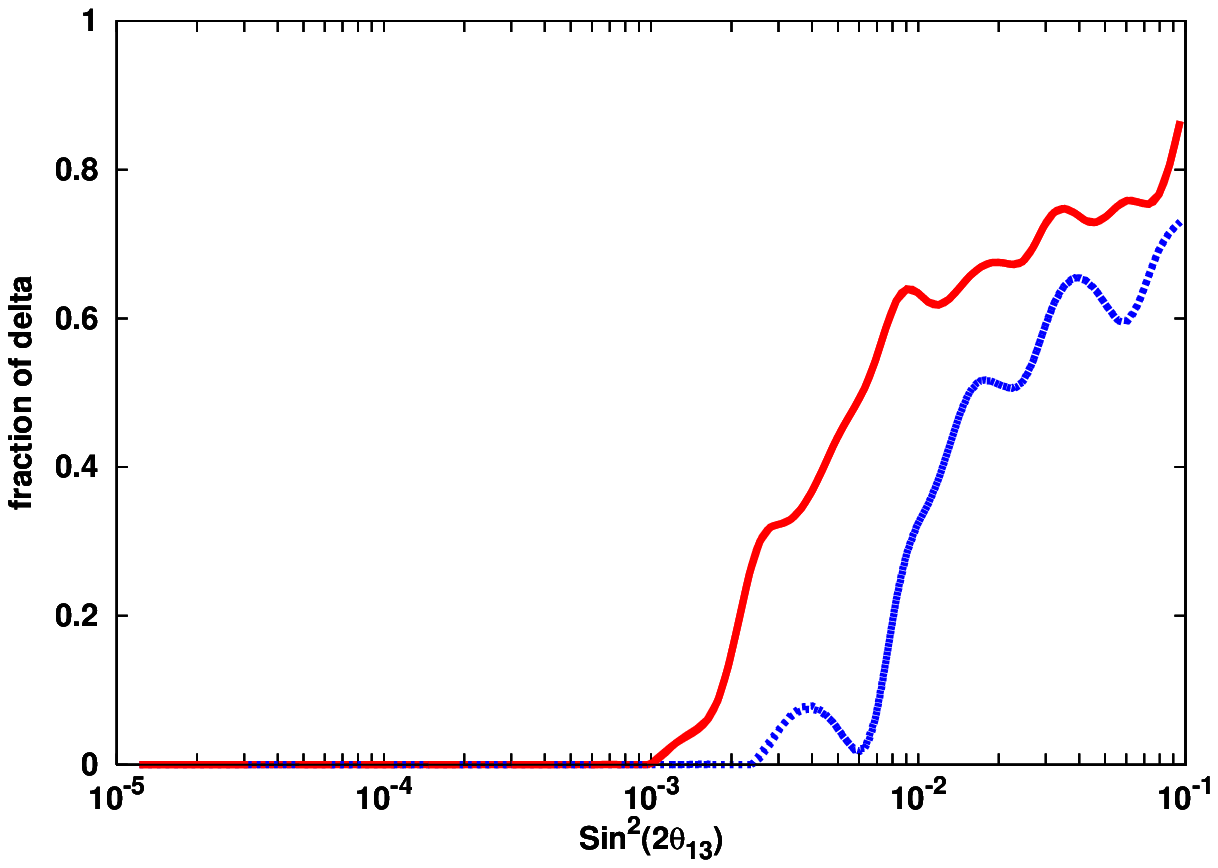}} 
\subfigure{\includegraphics[width=8cm]{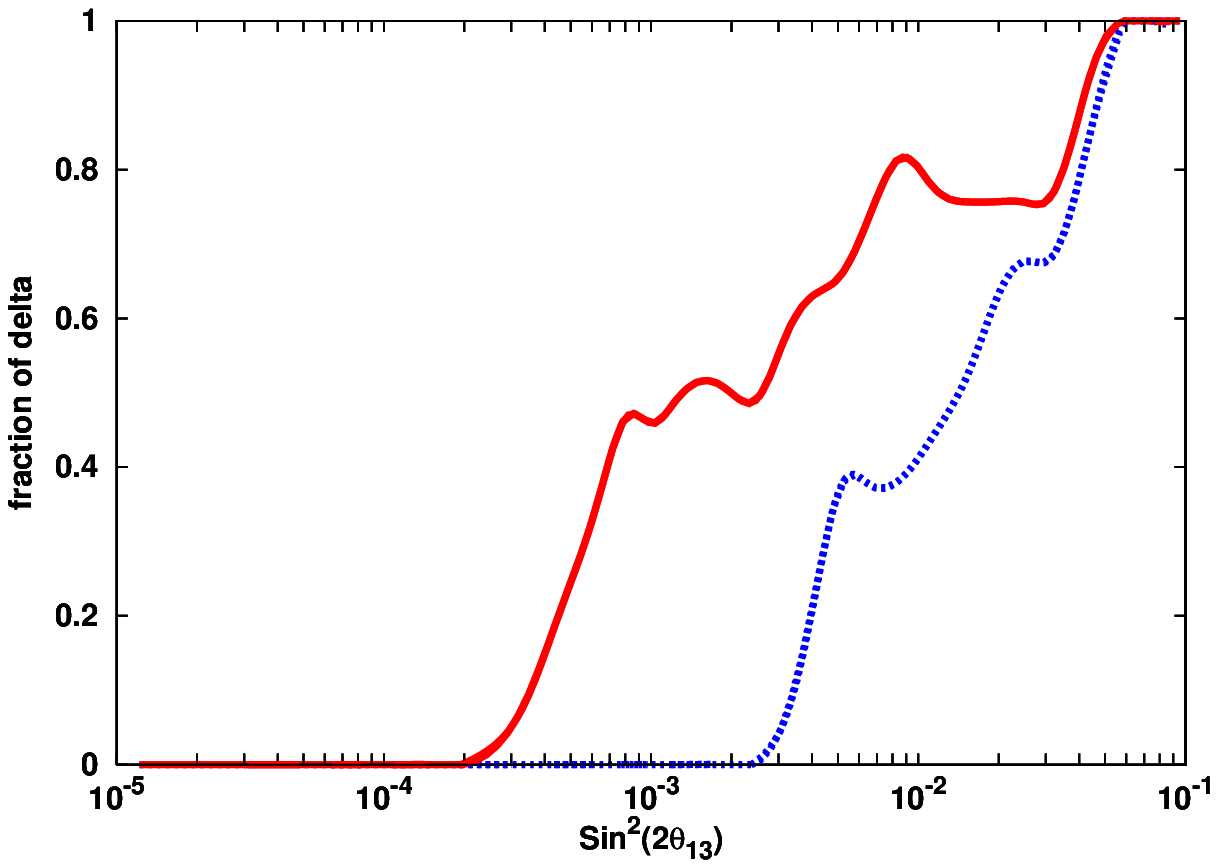}}\\
\end{center}
\caption{Same as Fig.~\ref{Fi:hierarchyWC}, but with negligible error
  in the value of the assumed neutrino parameters.}
\label{Fi:hierarchynoprior}
\end{figure}

On the other hand, we show in Fig.~\ref{Fi:hierarchynoprior} the
extent to which mass hierarchy determination could improve with better
knowledge of the assumed neutrino oscillation parameters. Again, we
only show the case of setup III-WC (left panel) and setup IV-WC (right
panel) with the same designations as previous, i.e., blue dashed lines
for the beta-beam only contribution and red solid lines for the
performance of the total flux (i.e., adding the electron capture flux).
The difference between Figs.~\ref{Fi:hierarchyWC}
and~\ref{Fi:hierarchynoprior} is substantial. While the qualitative 
behaviour of the relative contribution of the beta-beam only part is
very similar, in the case of perfect knowledge of the assumed
parameters, the determination of the neutrino mass hierarchy is
possible for 50\% of the values of $\delta$ down to $\sin^2
2\theta_{13} \simeq 8 \times 10^{-4}$ for setup IV-WC (right panel of
Fig.~\ref{Fi:hierarchynoprior}). This improvement represents about an 
order of magnitude with respect to the case depicted in the right
panel of Fig.~\ref{Fi:hierarchyWC}.

\section{Summary and conclusions}
\label{sec:conclusions}

Determining the value of $\theta_{13}$, the type of neutrino mass
ordering and the presence of leptonic CP-violation will be one of the
main goals of the future experimental neutrino program which is under
intensive discussion at present. In the present article we have
studied a new type of experimental setup which combines a beta-beam
with an electron capture beam. This can be achieved naturally by using
nuclei which can decay into both channels. We have studied this idea
using the nuclide $^{156}$Yb which has favourable beta-decay and
electron capture branching ratios, and only a small alpha decay
contribution. This combination is very powerful as the EC channel
provides a high energy signal at a well known energy, while the
beta-beam provides coverage of the first and second oscillation
maxima. The allowed regions in the ($\theta_{13}$, $\delta$) plane for
the two separate channels have a limited overlap resulting in a good
resolution of the intrinsic degeneracy. We have understood the main
features of this synergy by an analytical study of the oscillation
probability. It should be stressed that this setup does not require
two polarities but reaches a very good sensitivity by only using the
neutrino channel through full exploitation of the oscillatory pattern
of the appearance probability.
 
We have performed a detailed study of the dependence of the physics
reach of this experimental technique by considering six different
setups: two values for the ion boost factor $\gamma = 166$ and $369$;
two choices for the detector: a 50~kton TASD or LiAr and a 0.5~Mton WC
detector; and three baselines: CERN-Frejus, CERN-Canfranc,
CERN-Boulby. This allowed us to study the impact of the count rate, 
choice of baseline and the tuning of the energy of the beta-beam and EC
beam to the oscillatory pattern. We find that the setups with low
gamma and 50~kton detectors have very poor physics reach, owing to the
limited event rates. The information on CP-violation is mainly provided
by the high energy EC signal. We studied the options with $\gamma =
369$, the highest value of the boost factor allowed by an upgraded SPS.
Setups III and III-WC, which use the CERN-Canfranc baseline, have
larger count rates and a better tuning of the beam to the oscillatory
pattern, with respect to their CERN-Boulby counterparts: setups IV and
IV-WC. This results in a very good ability to measure the parameters,
see Fig.~\ref{Fi:CPWC}. In particular these setups provide the best
sensitivity to CP-violation for positive values of $\delta$. However,
for negative $\delta$, owing to the relatively short distance, the
type of hierarchy can be resolved only for very large values of
$\theta_{13}$. The sign-degeneracy prevents discovery of CP-violation
in this case, see Fig.~\ref{Fi:CC369clone}. The CERN-Boulby setups, IV
and IV-WC, suffer from smaller count rates and poor tuning of the
beta-beam to the oscillation pattern. However, they provide a much
better determination of the hierarchy and possess a good reach to
CP-violation for $\delta<0$, even if the mass ordering is not
known. Comparing the two baseline options, if the hierarchy is known
to be normal from other neutrino experiments, the CERN-Canfranc option
has an improved physics reach, while if the ordering is not known, the
CERN-Boulby baseline outperforms the shorter option. For the high
statistics scenario, one gets sensitivity to CP-violation down to
values of $\sin^{2}2\theta_{13} \sim 3\times10^{-5}$ at 99\% CL for a
WC detector at Canfranc, and $\sin^{2}2\theta_{13} \sim 10^{-4}$ for a
WC detector at Boulby.

In conclusion, we have presented the novel idea of using a single beam
which combines neutrinos from beta and electron capture decays and have
demonstrated the physics reach of several possible setups. We have
shown that the combination of these two types of beams achieves
remarkable results. This could naturally be done with the use 
$^{156}$Yb, which has comparable beta-beam and electron capture
branching ratios. As both beams are produced from a single isotope,
this combination cannot be further optimised. An analogous setup would
be obtained if a beta-beam and an electron capture beam sourced from
different ions are combined. In this case, further optimisation of the 
experiment would be allowed, for suitable choices of baselines,
Lorentz boost factors, detector size and technology, possibly
achieving an even better physics reach than the one found in the
present study.


\acknowledgments

We would like to thank O.~Mena and W.~Winter for useful discussions.
CE acknowledges the support from the Spanish Generalitat Valenciana
and thanks the IPPP for hospitality at the initial stage of this
study. CO acknowledges the support of a STFC studentship and overseas
fieldwork support. In addition, CO and SPR would like to thank the
Department de Fisica Te\`{o}rica, Universitat de Val\`encia for
hospitality during the middle stages of this work. SPR is supported by 
the Portuguese FCT through the projects POCI/FP/81919/2007,
CERN/FP/83503/2008 and CFTP-FCT UNIT 777, which are partially funded
through POCTI (FEDER). SPR is also partially supported by the Spanish
Grant FPA2005-01678 of the MCT. CO and SP acknowledge the support of
CARE, contract number RII3-CT-2003-506395.



\end{document}